\documentclass[preprint,12pt]{elsarticle}

\biboptions{numbers,sort&compress}

\usepackage{amssymb}
\usepackage{amsmath}
\usepackage{graphicx}
\usepackage{newtxtext}
\usepackage{newtxmath}
\usepackage{array}
\usepackage{subfigure}
\usepackage{enumitem}
\usepackage{tabularx}
\usepackage{booktabs}

\usepackage{latexsym,bm}
\usepackage{hyperref}

\usepackage{amsmath}
\usepackage{algorithm}
\usepackage{algpseudocode}
\usepackage{amsfonts}

\begin{document}

\bibliographystyle{elsarticle-num}

\begin{frontmatter}



\title{Enhancing generalizability of machine learning general effective-viscosity turbulence model via tensor basis normalization}


\author[1,2]{Ziqi Ji}
\author[2]{Penghao Duan}
\author[1]{Gang Du}
\affiliation[1]{organization={School of Energy and Power Engineering},
            addressline={Beihang University}, 
            city={Beijing},
            postcode={100191}, 
            country={China}}
\affiliation[2]{organization={Department of Mechanical Engineering},
            addressline={City University of Hong Kong}, 
            city={Hong Kong},
            postcode={999077}, 
            country={China}}

\begin{abstract}
With the rapid advancement of machine learning techniques, the development and study of machine learning turbulence models have become increasingly prevalent. As a critical component of turbulence modeling, the constitutive relationship between the Reynolds stress tensor and the mean flow quantities, modeled using machine learning methods, faces a pressing challenge: the lack of generalizability. To address this issue, we propose a novel tensor basis normalization technique to improve machine learning turbulence models, grounded in the general effective-viscosity hypothesis. In this study, we utilize direct numerical simulation (DNS) results of periodic hill flows as training data to develop a symbolic regression-based turbulence model based on the general effective-viscosity hypothesis. Furthermore, we construct a systematic validation dataset to evaluate the generalizability of our symbolic regression-based turbulence model. This validation set includes periodic hills with different aspect ratios from the training dataset, zero pressure gradient flat plate flows, three-dimensional incompressible flows over a NACA0012 airfoil, and transonic axial compressor rotor flows. These validation cases exhibit significant flow characteristics and geometrical variations, progressively increasing their differences from the training dataset. Such a diverse validation set is a robust benchmark to assess the generalizability of the proposed turbulence model. Finally, we demonstrate that our symbolic regression-based turbulence model performs effectively across validation cases, encompassing various separation features, geometries, and Reynolds numbers.
\end{abstract}

\begin{graphicalabstract}
\end{graphicalabstract}

\begin{highlights}
\item We propose a novel tensor basis normalization method to substantially improve the generalizability of machine learning turbulence models based on Pope's general effective-viscosity hypothesis.
\item We develop a symbolic regression strategy to ensure the results are insensitive to hyperparameter selection.
\end{highlights}

\begin{keyword}
Machine learning \sep Turbulence model \sep 
General effective-eddy hypothesis \sep Generalizability 



\end{keyword}

\end{frontmatter}




\section{Introduction}
\label{sec:Introduction}

Modern engineering design demands fast and high-fidelity computational fluid dynamics (CFD) simulation results. However, achieving fast and high-fidelity CFD simulations for complex engineering flows is challenging due to the high computational cost of large eddy simulation (LES) and DNS and the inaccuracy of traditional turbulence models in Reynolds-averaged Navier–Stokes (RANS) simulations. With the rapid development of machine learning, these methods are increasingly being applied to turbulence modeling. Many machine learning turbulence models have been proposed, yielding excellent results.

The uncertainties introduced by RANS turbulence models can be categorized into four physical levels \citep{duraisamy_turbulence_2019,he_turbo-oriented_2022}:

Physical level 1: The loss of fluctuation information due to the ensemble averaging of the RANS equations, which is an inevitable consequence of the RANS method.

Physical level 2: The uncertainty arising from the assumptions regarding the constitutive relationship between the Reynolds stress tensor and the mean flow quantities (for instance, the uncertainties brought about by the Boussinesq hypothesis).

Physical level 3: The uncertainty originating from the form of the turbulence model transport equation.

Physical level 4: The uncertainty associated with the calibration of the turbulence model coefficients.

Physical level 2 represents the most significant uncertainty physical level in RANS turbulence models and has attracted substantial attention in machine learning turbulence modeling research. The majority of machine learning turbulence model studies focusing on physical level 2 are founded upon the general effective-viscosity hypothesis proposed by Pope \cite{pope_more_1975}, which postulates that the non-dimensional Reynolds deviatoric tensor can be expressed as a finite tensor polynomial.

The groundbreaking advancements in deep learning techniques, particularly in computer vision and natural language processing, have significantly enhanced their potential for developing sophisticated machine learning turbulence models. Building upon several early-stage studies \cite{duraisamy2015new, tracey2015machine}, Ling et al. \cite{ling_reynolds_2016} pioneered a tensor basis neural network architecture with embedded Galilean invariance predicated on the general effective-viscosity hypothesis, thereby establishing a robust theoretical framework for deep learning applications in turbulence modeling. Subsequently, the scientific community has conducted extensive investigative research on neural network turbulence models, introducing numerous innovative architectural frameworks and achieving remarkable progress \citep{wu_physics-informed_2018, cruz_use_2019, yin_feature_2020, zhu_turbulence_2021, zhang_ensemble_2022, yin_iterative_2022, liu_learning_2023, zhang_physical_2023, shan_turbulence_2023, cai_revisiting_2024, ji_tensor_2024}. Nevertheless, the intrinsic ``black box'' characteristics inherent to neural network models have substantially limited their comprehensive application in elucidating the underlying physical mechanisms of turbulence and their broader generalizability across varied flow regimes.

Symbolic regression methods, with their distinctive advantage of generating explicit analytical expressions, demonstrate exceptional utility in mining complex functional relationships. Weatheritt and Sandberg \cite{weatheritt_novel_2016} pioneered the application of symbolic regression techniques to turbulence modeling based on the general effective-viscosity hypothesis, establishing an innovative framework for evolutionary algorithm-based symbolic regression turbulence models. Subsequently, the scientific community has conducted systematic investigations into the application of symbolic regression in turbulence modeling, proposing numerous innovative models that robustly validate the immense potential of symbolic regression in constructing explicit turbulence constitutive relationships \citep{zhao_rans_2020, xie_data-driven_2023, fang_toward_2023, lav_coupled_2023, wu_enhancing_2023, he_field_2024, wu_development_2024, li_evolutionary_2024, ji_interpreting_2025}. Nevertheless, the performance of these models regarding generalizability and physical interpretability remains subject to comprehensive evaluation and systematic verification.

Current machine learning turbulence models predicated on the general effective-viscosity hypothesis exhibit substantial limitations in generalization performance, demonstrating a notable disparity compared to corresponding research at the physical level 3. Wu et al. \cite{wu_development_2024} proposed a systematic, quantitative evaluation framework for assessing turbulence model generalization performance (as delineated in Table \ref{Table: Generalizaion_level_of_1_to_4}). Empirical evidence suggests that through innovative methodologies such as conditional field inversion \citep{wu_development_2024} and the rubber-band model \citep{bin_constrained_2024}, machine learning turbulence models operating at the physical level 3 have successfully attained generalization level 3. In marked contrast, physical level 2 machine learning turbulence models founded on the general effective-viscosity hypothesis continue to encounter formidable obstacles regarding generalization performance, with an overwhelming majority failing to satisfy even generalization level 2 criteria. This pronounced discrepancy underscores the necessity and urgency of developing novel methodological approaches to enhance the generalization capacities of physical level 2 machine learning turbulence models.

\begin{table}
    \centering

    \caption{Generalizaion level of 1 to 4 \citep{wu_development_2024}}
    
    \begin{tabular}{p{0.2\linewidth} p{0.3\linewidth} p{0.5\linewidth}}
    \hline
        Level & Definition & Example \\\hline
        
        Generalization level 1 & The model performs well in a series of geometries similar to the training set. & The model is trained on one periodic hill and can be generalized to other periodic hills with different aspect ratios.  \\ 
        
        Generalization level 2 & The model does not negatively affect the baseline model's accuracy in simple wall-attached flows. & The model is trained on some separated flow cases and can be as accurate as its baseline model on the zero-pressure-gradient flat plate. \\ 
        
        Generalization level 3 & The model performs well in test cases that have separation features similar to the training set but completely different geometries and Reynolds numbers. & (1) The model is trained on a case where separation is caused by a blunt geometry and is tested to be effective in other cases that have completely different blunt geometries and Reynolds numbers. 
        
        (2) The model is trained on a case where separation starts from a smooth surface (caused by a negative pressure gradient) and is tested to be effective on other smooth surface separations that have completely different geometries and Reynolds numbers. \\
        
        Generalization level 4 & The model performs well in a series of test cases with completely different separation features, geometries, and Reynolds numbers. & The model can accurately predict separation caused by blunt geometry as well as separation starting from a smooth surface. \\ \hline
    \end{tabular}

    \label{Table: Generalizaion_level_of_1_to_4}
    
\end{table}

In this work, we introduce an innovative tensor basis normalization technique and apply it to enhance the generalizability of machine learning turbulence models to generalization level 4. Our study utilizes DNS results from incompressible periodic hills \cite{xiao_flows_2019} as a benchmark. The assessment results demonstrate that turbulence models based on our tensor basis normalization technique preserve the law of the wall simulation performance compared to the baseline $k$-$\epsilon$ model while accurately simulating three-dimensional NACA0012 flows and transonic axial compressor rotor flows. The latter are particularly challenging as they are influenced by bluff-body separation and adverse pressure gradient separation while featuring extremely high rotation speeds and shock wave phenomena within the flow field. These transonic axial compressor rotor flows represent complex flow conditions that differ substantially from the training dataset and closely resemble engineering applications. Based on these results, we conclude that our machine learning turbulence models successfully achieve generalization level 4.

The structure of this paper is organized as follows: Section \ref{sec:Methodology} describes the methodology employed in this work. Section \ref{sec:Results} presents and analyzes the results. Section \ref{sec: Discussion} discusses the symbolic regression equation of the proposed turbulence model. Section \ref{sec: Conclusion} provides the conclusions of this study. \ref{sec: Detail_of_symbolic_regression results} details the training results of the symbolic regression. \ref{sec: Results_of_symbolic_regression-based_turbulence_model_without_normalization} evaluates the performance of the symbolic regression-based turbulence model without normalization.

\section{Methodology}\label{sec:Methodology}

\subsection{General effective-viscosity hypothesis and its normalized form}
\label{sec:General effective-viscosity hypothesis and its normalized form}

According to Pope's general effective-viscosity hypothesis \citep{pope_more_1975}, the non-dimensional Reynolds deviatoric tensor can be expressed as:

\begin{align}
\left\{\begin{array}{l}
\boldsymbol{b}=\sum_{i=1}^w g_i \boldsymbol{T}_i \\
g_i=f\left(I_1 \sim I_a, q_1 \sim q_b\right)
\end{array}\right.
\label{eq: general effective-viscosity hypothesis}
\end{align}
where $\bm{b}$ denotes the non-dimensional Reynolds deviatoric tensor, defined as $\boldsymbol{b}=\boldsymbol{\tau} / 2k-\boldsymbol{\mathrm{I}} / 3$, $\boldsymbol{\tau}$ is Reynolds stress. The terms $\left(I_1, \ldots, I_a\right)$ represent the tensor invariants, while $\left(q_1, \ldots, q_b\right)$ signify additional mean flow characteristics. $\{\boldsymbol{T}_i , i=1, \ldots, w\}$ constitute the tensor basis, which can be derived from the flow field results. The coefficients $g_i$ are functions of the tensor invariants $\left(I_1, \ldots, I_a\right)$ and the aforementioned mean flow characteristics $\left(q_1, \ldots, q_b\right)$.

Given that $\boldsymbol{T}_i$ constitute the tensor basis, the tensor basis coefficients $g_i$ can be derived through the following optimization:
\begin{align}
g_i=\underset{g_i}{\arg \min }\left(\left\|\boldsymbol{b}-g_i \boldsymbol{T}_i\right\|_F^2\right)=\frac{\boldsymbol{b}: \boldsymbol{T}_i}{\left\|\boldsymbol{T}_i\right\|_F^2},
\label{eq: gi}
\end{align}
where $||\cdot||_F$ represents the Frobenius norm, while ``$:$'' denotes the double contraction. 

However, the magnitude of $g_i$ calculated via Eq. (\ref{eq: gi}) can become exceedingly large as $\left\|\boldsymbol{T}_i\right\|_F$ approaches zero, presenting a significant challenge for machine learning algorithms in accurately predicting $g_i$. To address this issue, we propose a novel normalization technique for the tensor basis coefficients $g_i$, enhancing the predictive capabilities of machine learning methods. The normalized tensor basis coefficients are expressed as:
\begin{align}
\hat{g}_i=\frac{\boldsymbol{b}: \frac{\boldsymbol{T}_i}{\left\|\boldsymbol{T}_i\right\|_F}}{\left\|\frac{\boldsymbol{T}_i}{\left\|\boldsymbol{T}_i\right\|_F}\right\|_F^2}=\boldsymbol{b}: \frac{\boldsymbol{T}_i}{\left\|\boldsymbol{T}_i\right\|_F}.
\label{eq: gi_normalized}
\end{align}
Consequently, we define the normalized tensor basis as $\hat{\boldsymbol{T}}_i=\frac{\boldsymbol{T}_i}{\left\|\boldsymbol{T}_i\right\|_F}$. Accordingly, Eq. (\ref{eq: general effective-viscosity hypothesis}) can be reformulated as:
\begin{align}
\left\{\begin{array}{l}
\boldsymbol{b}=\sum_{i=1}^w \hat{g}_i \hat{\boldsymbol{T}}_i \\
\hat{g}_i=\hat{f}\left(I_1 \sim I_a, q_1 \sim q_b\right)
\end{array}\right.
\label{eq: gihat}
\end{align}
The relationship between the normalized tensor basis coefficients $\hat{g}_i$ and the original tensor basis coefficients $g_i$ can be mathematically expressed as follows:
\begin{align}
\hat{g}_i= \left\|\boldsymbol{T}_i\right\|_F g_i 
\label{eq: gihat_and_gi}
\end{align}

The non-dimensional Reynolds deviatoric tensor $\boldsymbol{b}$ is subject to the following physical realizability constraints:
\begin{align}
\left\{\begin{array}{ll}
-\frac{1}{3} \leq b_{i j} \leq \frac{2}{3} & \text { for } \mathrm{i}=\mathrm{j} \\[1mm]
-\frac{1}{2} \leq b_{i j} \leq \frac{1}{2} & \text { for } \mathrm{i} \neq \mathrm{j}
\end{array}\right.
\end{align}
Consequently, the values of $\hat{g}_i$ are inherently bounded, whereas the values of $g_i$ remain unbounded.

Concerning the general effective-viscosity hypothesis, the tensor basis $\{\boldsymbol{T}_i , i=1, \ldots, w\}$ constitutes a complete basis set wherein the directional components of each basis tensor carry greater significance than their respective magnitudes. Therefore, separating the magnitude and directional information of the tensor basis is advantageous. We can effectively decouple these two aspects by employing Eq. (\ref{eq: gi_normalized}) and (\ref{eq: gihat}), allowing us to utilize machine learning techniques specifically for predicting the magnitude information.

\subsection{Input features and tensor basis}
\label{sec:Input features and tensor basis}

Wu et al. \cite{wu_physics-informed_2018} proposed a functional relationship between the Reynolds stress and mean flow quantities:
\begin{align}
\boldsymbol{\tau}=g(\boldsymbol{S}, \boldsymbol{R}, \nabla p, \nabla k)
\label{eq: Reynolds stress relationship}
\end{align}
$\boldsymbol{S}=\frac{1}{2}\left(\nabla \boldsymbol{U}+(\nabla \boldsymbol{U})^{\mathrm{T}}\right)$ and $\boldsymbol{R}=\frac{1}{2}\left(\nabla \boldsymbol{U}-(\nabla \boldsymbol{U})^{\mathrm{T}}\right)$ are the strain-rate tensor and rotation-rate tensor, respectively, while $\nabla p, \nabla k$ represent the pressure gradient and turbulence kinetic energy gradient, respectively. We normalize these 4 mean flow quantities based on the normalization factors proposed by Yin et al. \cite{yin_iterative_2022}, resulting in the normalized raw input $\hat{\boldsymbol{\alpha}}$, as shown in Table \ref{Table: input normalization}. The normalization scheme is as follows:
\begin{align}
\hat{\boldsymbol{\alpha}}=\frac{\boldsymbol{\alpha}}{\|\boldsymbol{\alpha}\|+|\beta|}
\label{eq: normalization scheme}
\end{align}
In this formulation, $\|\cdot\|$ represents the tensor norm, with the F-norm specifically adopted in this study. $|\cdot|$ refers to the absolute value of a scalar.

The functional relationship Eq. (\ref{eq: Reynolds stress relationship}) is required to fulfill the condition of rotational invariance, which can be denoted as:
\begin{align}
\boldsymbol{Q} \boldsymbol{\tau} \boldsymbol{Q}^T=h\left(\boldsymbol{Q S} \boldsymbol{Q}^T, \boldsymbol{Q R} \boldsymbol{Q}^T, \boldsymbol{Q} \nabla p, \boldsymbol{Q} \nabla k\right)
\label{eq: rotational relationship}
\end{align}
the matrix $\bm{Q}$ herein is orthogonal. Rotational invariance can be ensured by constructing the minimal complete tensor basis of the input set $\hat{\boldsymbol{Q}}=\{\hat{\boldsymbol{S}}, \hat{\boldsymbol{R}}, \hat{\nabla} p, \hat{\nabla} k\}$.

Based upon the derivation by Wu et al. \cite{wu_physics-informed_2018}, a minimal integrity basis satisfying rotational invariance for the input set $\hat{\boldsymbol{Q}}=\{\hat{\boldsymbol{S}}, \hat{\boldsymbol{R}}, \hat{\nabla} p, \hat{\nabla} k\}$ consists of 47 traces of independent matrix products. These 47 independent matrix products are detailed in Table \ref{Table: Invariant basis}.

\begin{table}
    \centering
    
    \caption{The normalization of strain-rate tensor, rotation-rate tensor, pressure gradient, and turbulence kinetic energy gradient. $\omega$ represents the specific dissipation rate, while $k$ represents the turbulent kinetic energy}
    
    \begin{tabular}{>{\centering\arraybackslash}m{3cm}>{\centering\arraybackslash}m{3cm}>{\centering\arraybackslash}m{3cm}>{\centering\arraybackslash}m{3cm}}
    \hline
        Normalized raw input $\hat{\bm{\alpha}}$ & Description & Raw input $\bm{\alpha}$ & Normalization factor $\beta$  \\\hline
        $\hat{\boldsymbol{S}}$ & Strain-rate tensor & $\boldsymbol{S}$ & $\omega$  \\ 
        $\hat{\boldsymbol{R}}$ & Rotation-rate tensor & $\boldsymbol{R}$ & $\omega$  \\ 
        $\hat{\nabla p}$ & Pressure gradient & $\nabla p$ & $\omega \sqrt{k}$ \\ 
        $\hat{\nabla k}$ & Turbulence kinetic energy gradient & $\nabla k$ & $\omega \sqrt{k}$ \\ \hline
    \end{tabular}
    
    \label{Table: input normalization}
\end{table}

\begin{table}
    \centering

    \caption{Minimal integrity basis for input set $\hat{\boldsymbol{Q}}=\{\hat{\boldsymbol{S}}, \hat{\boldsymbol{R}}, \hat{\nabla} p, \hat{\nabla} k\}$. The asterisk (*) denotes the comprehensive set of terms formed by cyclical permutations marked with antisymmetric tensor indicators, where, for instance, $\hat{\boldsymbol{R}}^2 \hat{\boldsymbol{A}}_p \hat{\boldsymbol{S}}^*$ represents both $\hat{\boldsymbol{R}}^2 \hat{\boldsymbol{A}}_p \hat{\boldsymbol{S}}$ and $\hat{\boldsymbol{A}}_p \hat{\boldsymbol{R}}^2 \hat{\boldsymbol{S}}$. The terms $\hat{\boldsymbol{A}}_p$ and $\hat{\boldsymbol{A}}_k$ are defined by the expressions $\hat{\boldsymbol{A}}_p = -\boldsymbol{I} \times \hat{\nabla p}$ and $\hat{\boldsymbol{A}}_k = -\boldsymbol{I} \times \hat{\nabla k}$, respectively.}
    
    \begin{tabular}{c>{\centering\arraybackslash}p{110mm}}
    \hline
        Feature index & Invariant basis  \\ \hline 
        1-2 & $\hat{\boldsymbol{S}}^2, \hat{\boldsymbol{S}}^3$ \\ 
        3-5 & $\hat{\boldsymbol{R}}^2, \hat{\boldsymbol{A}}_p^2, \hat{\boldsymbol{A}}_k^2$ \\ 
        6-14 & $\hat{\boldsymbol{R}}^2 \hat{\boldsymbol{S}}, \hat{\boldsymbol{R}}^2 \hat{\boldsymbol{S}}^2, \hat{\boldsymbol{R}}^2 \hat{\boldsymbol{S}} \hat{\boldsymbol{R}} \hat{\boldsymbol{S}}^2$, $\hat{\boldsymbol{A}}_p^2 \hat{\boldsymbol{S}}, \hat{\boldsymbol{A}}_p^2 \hat{\boldsymbol{S}}^2, \hat{\boldsymbol{A}}_p^2 \hat{\boldsymbol{S}} \hat{\boldsymbol{A}}_p \hat{\boldsymbol{S}}^2$, $\hat{\boldsymbol{A}}_k^2 \hat{\boldsymbol{S}}, \hat{\boldsymbol{A}}_k^2 \hat{\boldsymbol{S}}^2, \hat{\boldsymbol{A}}_k^2 \hat{\boldsymbol{S}} \hat{\boldsymbol{A}}_k \hat{\boldsymbol{S}}^2$ \\
        15-17 & $\hat{\boldsymbol{R}} \hat{\boldsymbol{A}}_p, \hat{\boldsymbol{A}}_p \hat{\boldsymbol{A}}_k, \hat{\boldsymbol{R}} \hat{\boldsymbol{A}}_k$ \\ 
        18-41 & $\hat{\boldsymbol{R}} \hat{\boldsymbol{A}}_p \hat{\boldsymbol{S}}, \hat{\boldsymbol{R}} \hat{\boldsymbol{A}}_p \hat{\boldsymbol{S}}^2, \hat{\boldsymbol{R}}^2 \hat{\boldsymbol{A}}_p \hat{\boldsymbol{S}}^*, \hat{\boldsymbol{R}}^2 \hat{\boldsymbol{A}}_p \hat{\boldsymbol{S}}^{2^*}, \hat{\boldsymbol{R}}^2 \hat{\boldsymbol{S}} \hat{\boldsymbol{A}}_p \hat{\boldsymbol{S}}^{2^*}$, $\hat{\boldsymbol{R}} \hat{\boldsymbol{A}}_k \hat{\boldsymbol{S}}, \hat{\boldsymbol{R}} \hat{\boldsymbol{A}}_k \hat{\boldsymbol{S}}^2, \hat{\boldsymbol{R}}^2 \hat{\boldsymbol{A}}_k \hat{\boldsymbol{S}}^*, \hat{\boldsymbol{R}}^2 \hat{\boldsymbol{A}}_k \hat{\boldsymbol{S}}^{2 *}, \hat{\boldsymbol{R}}^2 \hat{\boldsymbol{S}} \hat{\boldsymbol{A}}_k \hat{\boldsymbol{S}}^{2^*}$, $\hat{\boldsymbol{A}}_p \hat{\boldsymbol{A}}_k \hat{\boldsymbol{S}}, \hat{\boldsymbol{A}}_p \hat{\boldsymbol{A}}_k \hat{\boldsymbol{S}}^2, \hat{\boldsymbol{A}}_p^2 \hat{\boldsymbol{A}}_k \hat{\boldsymbol{S}}^*, \hat{\boldsymbol{A}}_p^2 \hat{\boldsymbol{A}}_k \hat{\boldsymbol{S}}^{2^*}, \hat{\boldsymbol{A}}_p^2 \hat{\boldsymbol{S}}_k \hat{\boldsymbol{A}}_k \hat{\boldsymbol{S}}^{2^*}$ \\ 
        42 & $\hat{\boldsymbol{R}} \hat{\boldsymbol{A}}_p \hat{\boldsymbol{A}}_k$ \\ 
        43-47 & $\hat{\boldsymbol{R}} \hat{\boldsymbol{A}}_p \hat{\boldsymbol{A}}_k \hat{\boldsymbol{S}}, \hat{\boldsymbol{R}} \hat{\boldsymbol{A}}_k \hat{\boldsymbol{A}}_p \hat{\boldsymbol{S}}, \hat{\boldsymbol{R}} \hat{\boldsymbol{A}}_p \hat{\boldsymbol{A}}_k \hat{\boldsymbol{S}}^2, \hat{\boldsymbol{R}} \hat{\boldsymbol{A}}_k \hat{\boldsymbol{A}}_p \hat{\boldsymbol{S}}^2, \hat{\boldsymbol{R}} \hat{\boldsymbol{A}}_p \hat{\boldsymbol{S}} \hat{\boldsymbol{A}}_k \hat{\boldsymbol{S}}^2$ \\ \hline
    \end{tabular}

    \label{Table: Invariant basis}
\end{table}

Owing to the small absolute values and unsmooth distribution of high-degree invariants, reducing their usage can contribute to enhanced robustness and increased computational speed \citep{yin_iterative_2022}. The tensor invariants are categorized according to degree \citep{ji_tensor_2024}, as in Table \ref{Table: degree}.

We adopt the tensor basis proposed by Pope \cite{pope_more_1975}. The tensor basis, $\{\boldsymbol{T}_1$ $\sim$$\boldsymbol{T}_{10}\}$, are as follows:
\begin{equation}
\begin{array}{l}
\left.\begin{array}{ll}
\boldsymbol{T}_1=\hat{\boldsymbol{S}} & \boldsymbol{T}_6=\hat{\boldsymbol{R}}^2 \hat{\boldsymbol{S}}+\hat{\boldsymbol{S}} \hat{\boldsymbol{R}}^2-\frac{2}{3} \boldsymbol{I} \cdot \operatorname{Tr}\left(\hat{\boldsymbol{S}} \hat{\boldsymbol{R}}^2\right) \\ [2mm]
\boldsymbol{T}_2=\hat{\boldsymbol{S}} \hat{\boldsymbol{R}}-\hat{\boldsymbol{R}} \hat{\boldsymbol{S}} & \boldsymbol{T}_7=\hat{\boldsymbol{R}} \hat{\boldsymbol{S}} \hat{\boldsymbol{R}}^2-\hat{\boldsymbol{R}}^2 \hat{\boldsymbol{S}} \hat{\boldsymbol{R}} \\ [2mm]
\boldsymbol{T}_3=\hat{\boldsymbol{S}}^2-\frac{1}{3} \boldsymbol{I} \cdot \operatorname{Tr}\left(\hat{\boldsymbol{S}}^2\right) & \boldsymbol{T}_8=\hat{\boldsymbol{S}} \hat{\boldsymbol{R}} \hat{\boldsymbol{S}}^2-\hat{\boldsymbol{S}}^2 \hat{\boldsymbol{R}} \hat{\boldsymbol{S}} \\ [2mm]
\boldsymbol{T}_4=\hat{\boldsymbol{R}}^2-\frac{1}{3} \boldsymbol{I} \cdot \operatorname{Tr}\left(\hat{\boldsymbol{R}}^2\right) & \boldsymbol{T}_9=\hat{\boldsymbol{R}}^2 \hat{\boldsymbol{S}}^2+\hat{\boldsymbol{S}}^2 \hat{\boldsymbol{R}}^2-\frac{2}{3} \boldsymbol{I} \cdot \operatorname{Tr}\left(\hat{\boldsymbol{S}}^2 \hat{\boldsymbol{R}}^2\right) \\ [2mm]
\boldsymbol{T}_5=\hat{\boldsymbol{R}} \hat{\boldsymbol{S}}^2-\hat{\boldsymbol{S}}^2 \hat{\boldsymbol{R}} & \boldsymbol{T}_{10}=\hat{\boldsymbol{R}} \hat{\boldsymbol{S}}^2 \hat{\boldsymbol{R}}^2-\hat{\boldsymbol{R}}^2 \hat{\boldsymbol{S}}^2 \hat{\boldsymbol{R}}
\end{array}\right\}
\end{array}
\label{eq: tensor basis}
\end{equation}
where the symbol $\boldsymbol{I}$ denotes the third-order identity matrix, while Tr($\cdot$) signifies the trace of a matrix. Extra features are enumerated in Table \ref{Table:extra features}. The following formula was employed for the non-dimensionalization of additional features:
\begin{align}
q_\beta=\frac{\hat{q}_\beta}{\left|\hat{q}_\beta\right|+\left|q_\beta^*\right|}
\label{eq: extra features non-dimensionalization}
\end{align}
except for $q_3$, $q_9$, and $q_{12}$ because they are inherently dimensionless and therefore does not require non-dimensionalization.

\begin{table}
    \centering

    \caption{Normalized tensor basis and tensor invariants classified by degree}
    
    \begin{tabular}{cc>{\centering\arraybackslash}p{50mm}}
    \hline
        Degree & Normalized tensor basis & Invariants  \\ \hline
        1 & $\boldsymbol{\hat{T}}_1$ & -  \\ 
        2 & $\boldsymbol{\hat{T}}_2$, $\boldsymbol{\hat{T}}_3$, $\boldsymbol{\hat{T}}_4$ & $I_1$, $I_3$, $I_4$, $I_5$, $I_{15}$, $I_{16}$, $I_{17}$  \\ 
        3 & $\boldsymbol{\hat{T}}_5$, $\boldsymbol{\hat{T}}_6$ & $I_2$, $I_6$, $I_9$, $I_{12}$, $I_{18}$, $I_{26}$, $I_{34}$, $I_{42}$  \\ 
        4 & $\boldsymbol{\hat{T}}_7$, $\boldsymbol{\hat{T}}_8$, $\boldsymbol{\hat{T}}_9$ & $I_7$, $I_{10}$, $I_{13}$, $I_{19}$, $I_{20}$, $I_{21}$, $I_{27}$, $I_{28}$, $I_{29}$, $I_{35}$, $I_{36}$, $I_{37}$, $I_{43}$, $I_{44}$  \\ 
        5 & $\boldsymbol{\hat{T}}_{10}$ & $I_{22}$, $I_{23}$, $I_{30}$, $I_{31}$, $I_{38}$, $I_{39}$, $I_{45}$, $I_{46}$  \\ 
        6 & - & $I_{8}$, $I_{11}$, $I_{14}$, $I_{24}$, $I_{25}$, $I_{32}$, $I_{33}$, $I_{40}$, $I_{41}$, $I_{47}$ \\ \hline
    \end{tabular}
    
    \label{Table: degree}
\end{table}

\begin{table}
    \centering

    \caption{List of extra features. $\boldsymbol{U}$ refers to the mean velocity, $d$ denotes the distance to the nearest wall, $\nu$ is the kinematic viscosity, $p$ symbolizes the mean pressure, $\varepsilon$ is the turbulence dissipation rate, $\rho$ is the mean density, and $c$ stands for the local speed of sound. $\boldsymbol{\Omega}=\nabla \times \boldsymbol{U}$ is vorticity. The dimensionless parameter $r_d=\left(v_t+v\right) / \kappa^2 d^2 \sqrt{U_{i j} U_{i j}}$ serves as a metric for quantifying the distance from the wall, where $\kappa$ is assigned a value of 0.41, commonly known as the von Kármán constant}
    
    \begin{tabular}{>{\centering\arraybackslash}p{20mm}>{\centering\arraybackslash}p{45mm}>{\centering\arraybackslash}p{30mm}>{\centering\arraybackslash}p{25mm}}
    \hline
        Normalized extra features $q_\beta$ & Description & Origin extra features $\hat{q}_\beta$ & Normalization factor $q_\beta^*$ \\ \hline 
        $q_1$ & Ratio of excess rotation rate to strain rate & $\frac{1}{2}\left(\|\boldsymbol{R}\|^2_F-\|\boldsymbol{S}\|^2_F\right)$ & $\|\boldsymbol{S}\|^2_F$ \\ 
        $q_2$ & Ratio of turbulent/mean kinetic energy & $k$ & $\frac{1}{2} U_i U_i$ \\ 
        $q_3$ & Wall-distance based Reynolds number & $\min \left(\frac{\sqrt{k} d}{50 \nu}, 2\right)$ & Not applicable  \\ 
        $q_4$ & Pressure gradient along streamline & $U_k \frac{\partial P}{\partial x_k}$ & $\frac{d^2}{\|\boldsymbol{S}\|^3_F}$ \\ 
        $q_5$ & Ratio of turbulent time scale to mean strain time scale & $\frac{k}{\varepsilon}$ & $\frac{1}{\|\boldsymbol{S}\|_F}$ \\ 
        $q_6$ & Ratio of pressure normal stresses to shear stresses & $\sqrt{\frac{\partial P}{\partial x_i} \frac{\partial P}{\partial x_i}}$ & $\frac{1}{2} \rho \frac{\partial U_k^2}{\partial x_k}$ \\ 
        $q_7$ & Nonorthogonality between velocity and its gradient & $\left|U_i U_j \frac{\partial U_i}{\partial x_j}\right|$ & $\epsilon$ \\ 
        $q_8$ & Turbulent Mach number & $\sqrt{k}$ & $c$  \\ 
        $q_9$ & Value of relative helicity density & $\frac{\boldsymbol{U} \cdot \boldsymbol{\Omega}}{|\boldsymbol{U}| \cdot|\boldsymbol{\Omega}|}$ & Not applicable  \\ 
        $q_{10}$ & turbulent kinetic energy & $k$ & $\nu\|\boldsymbol{S}\|_F$ \\ 
        $q_{11}$ & Ratio of turbulent time scale to mean strain time scale & $\|\boldsymbol{S}\|_F$ & $\omega$  \\ 
        $q_{12}$ & Marker of boundary layer & $1-\tanh \left(\left[8 r_d\right]^3\right)$ & Not applicable  \\ 
        $q_{13}$ & Ratio of turbulent/mean viscosity & $\nu_t$ & $\nu$ \\ \hline
    \end{tabular}

    \label{Table:extra features}
\end{table}

Based on Tables \ref{Table: Invariant basis} and \ref{Table:extra features}, we have identified 60 input features. This substantial number of features poses significant challenges for symbolic regression algorithms. Consequently, not all of these input features are utilized in our subsequent analyses. The methodology for feature selection is informed by the characteristics of the training dataset and will be discussed in detail in Section \ref{sec:Symbolic_regression_training_results}.

\subsection{Baseline model}
\label{sec:Baseline model}

The baseline model we used in this work is $k$-$\epsilon$ model proposed by Launder and Spalding \cite{launder1974application}. The governing equation of baseline model can be expressed by:
\begin{align}
\left\{\begin{aligned}
\frac{\partial \rho k}{\partial t}+\nabla \cdot(\rho \boldsymbol{U} k)-\nabla \cdot\left(\rho D_k \nabla k\right) &= \rho G-\frac{2}{3} \rho(\nabla \cdot \boldsymbol{U}) k-\rho \frac{\varepsilon+D}{k} k+S_k \\
\frac{\partial \rho \varepsilon}{\partial t}+\nabla \cdot(\rho \boldsymbol{U} \varepsilon)-\nabla \cdot\left(\rho D_{\varepsilon} \nabla \varepsilon\right) &= C_1 \rho G \frac{\varepsilon}{k}-\left(\left(\frac{2}{3} C_1-C_3\right) \rho(\nabla \cdot \boldsymbol{U}) \varepsilon\right) \\
& -C_2 f_2 \rho \frac{\varepsilon}{k} \varepsilon+\rho E+S_{\varepsilon}
\end{aligned}\right.
\label{eq: k-epsilon}
\end{align}
where $G=\nu_t\left(\nabla \boldsymbol{U}+\nabla \boldsymbol{U}^T-\tfrac{2}{3}(\nabla \cdot \boldsymbol{U}) \mathbf{I}\right): \nabla \boldsymbol{U}$ represents the turbulence production term, $E=2 \nu \nu_t|\nabla \nabla \boldsymbol{U}|^2$ denotes the wall-reflection term, and $f_2=1-0.3 \exp \left(-\min \left(\smash{\tfrac{k^4}{\nu^2 \varepsilon^2}}, 50\right)\right)$ is a damping function. Additionally, $D=2 \nu|\nabla(\sqrt{k})|^2$ signifies the diffusion term, $f_\mu=\exp \left({-3.4 / \left(1+{k^2 \over 50 \nu \varepsilon}\right)^2}\right)$ is another damping function, and $\nu_t=C_\mu f_\mu \tfrac{k^2}{\varepsilon}$ represents the eddy viscosity. The diffusion coefficients are given by $D_k=\tfrac{\nu_t}{\sigma_k}+\nu$ and $D_{\varepsilon}=\tfrac{\nu_t}{\sigma_{\varepsilon}}+\nu$ for the $k$ and $\varepsilon$ equations, respectively. The specific dissipation rate $\omega$, which is employed in Section \ref{sec:Input features and tensor basis} of this study, can be calculated using the following expression:
\begin{align}
\omega=\frac{\varepsilon}{C_\mu k},
\label{eq: omega}
\end{align}
where $C_\mu = 0.09$. Other empirical coefficients can be referenced in Launder and Spalding \cite{launder1974application}. 

\subsection{Symbolic regression}
\label{sec:Symbolic regression}

This work uses open-source multi-population evolutionary symbolic regression library PySR \citep{cranmer2023interpretablemachinelearningscience}. In our framework, the training target of symbolic regression are normalized coefficients $\hat{g}_i$. The primary hyperparameters of the symbolic regression algorithm are detailed in Table \ref{Table: Symbolic regression}. Within our framework, we have incorporated six mathematical operators (+, -, $\times$, $\div$, neg, inv). The unary operators ``neg'' and ``inv'' represent negation and inverse operations fundamentally distinct from their binary counterparts - and $\div$. Our training framework consists of 16 populations, each with 200 individuals. We perform 500 mutations for every 10 samples from each population per iteration. The maximum allowable complexity for any equation is set at 20, with a maximum depth of 10. Other hyperparameters are configured to the default settings of PySR.

\begin{table}
    \centering

    \caption{Main hyperparameters of the symbolic regression algorithm}
    
    \begin{tabular}{p{0.4\linewidth} p{0.4\linewidth}}
    \hline
        Principal parameters of the algorithm & Parameters setup  \\ \hline 
        Variables & Tensor invariants and extra features  \\ 
        Operators & +, -, $\times$, $\div$, neg, inv  \\ 
        Complexity of constants & 2  \\ 
        Number of populations & 16  \\ 
        Number of individuals in each population & 200  \\ 
        Number of total mutations to run, per 10 samples of the population, per iteration & 500  \\ 
        Max complexity of an equation & 20  \\ 
        Max depth of an equation & 10 \\ \hline
    \end{tabular}

    \label{Table: Symbolic regression}
    
\end{table}

To ensure the convergence of the symbolic regression, we have designed a robust algorithm that mitigates sensitivity to hyperparameter selection while guaranteeing convergence to solutions with minimal loss. The algorithmic procedure is outlined below, with the corresponding pseudocode presented in Algorithm \ref{alg:Symbolic_regression_training_strategy}.

1. Establishing an exemplary baseline hyperparameter configuration for symbolic regression algorithms.

2. Employing this baseline hyperparameter configuration for a statistically sufficient number $n$ of independent executions to identify all convergent equations of equivalent complexity within the established parameter space.

3. Double the training iterations and conduct $n$ independent executions to identify all convergent equations of equivalent complexity within this new parameter space.

4. Iteratively repeat step 3 until all convergent equations of equivalent complexity remain consistent across three consecutive training iteration increments. The convergent equation exhibiting the minimum loss function value constitutes the optimal solution for our requirements.

5. Evaluate the equation obtained in step 4; if its performance, despite exhibiting the minimum loss value, fails to meet the desired specifications, systematically examine convergent equations with progressively higher loss values until either a satisfactory solution is identified or all available convergent equations are deemed inadequate for the intended application.

\begin{algorithm}
\caption{Symbolic regression training strategy}
\label{alg:Symbolic_regression_training_strategy}
\begin{algorithmic}[1]
\State \textbf{Input:} Baseline hyperparameter configuration $\mathcal{H}$, performance $\mathcal{P}$, maximum training iterations $T_{\text{max}}$, statistical sample size $n$
\State \textbf{Output:} Optimal equation $\mathcal{E}_{\text{optimal}}$ 

\State Initialize $\mathcal{H}_{\text{baseline}} \gets \mathcal{H}$
\State Initialize $T \gets T_{\text{base}}$ (initial training iterations)
\State Initialize $\mathcal{E}_{\text{convergent}} \gets \emptyset$

\While{$T \leq T_{\text{max}}$}
    \State Perform $n$ independent executions of the symbolic regression algorithm using $\mathcal{H}_{\text{baseline}}$ and $T$ as the training iterations.
    \State Collect all convergent equations $\mathcal{E}_{T}$ of equivalent complexity in the current parameter space.
    
    \If{$\mathcal{E}_{T} = \mathcal{E}_{T-1} = \mathcal{E}_{T-2}$ (consistency across three iterations)}
        \State $\mathcal{E}_{\text{optimal}} \gets \arg\min_{\mathcal{E} \in \mathcal{E}_{T}} \text{Loss}(\mathcal{E})$
        \State \textbf{break}
    \Else
        \State $T \gets 2T$ (double the training iterations)
    \EndIf
\EndWhile

\If{$\mathcal{P}(\mathcal{E}_{\text{optimal}}) < \mathcal{P}_{\text{desired}}$}
    \For{each $\mathcal{E} \in \mathcal{E}_{\text{convergent}}$, sorted by increasing loss}
        \If{$\mathcal{P}(\mathcal{E}) \geq \mathcal{P}_{\text{desired}}$}
            \State $\mathcal{E}_{\text{optimal}} \gets \mathcal{E}$
            \State \textbf{break}
        \EndIf
    \EndFor
\EndIf

\State \textbf{Return} $\mathcal{E}_{\text{optimal}}$

\end{algorithmic}
\end{algorithm}

For instance, we selected a baseline hyperparameter configuration as presented in Table \ref{Table: Symbolic regression} and initially executed 50 iterations. Upon implementing the algorithmic procedure, we observed that all convergent equations of equivalent complexity remained consistent across 50, 100, and 200 iterations, demonstrating robust convergence behavior. Subsequently, we evaluated the equation with minimum loss value but found its performance insufficient for our requirements. Therefore, we systematically examined equations with progressively higher loss values until we identified one that satisfied our performance criteria. This final selected equation demonstrated hyperparameter insensitivity and met our performance demands, thus confirming its reliability for practical implementation.

\subsection{Framework}
\label{sec:Framework}

In this study, we employ TCAE as our CFD software for conducting simulations. TCAE, developed as an extension of the open-source CFD platform OpenFOAM, features a proprietary solver called ``redSolver'' that demonstrates exceptional robustness when modeling compressible and transonic flow regimes. In our investigation, this solver is specifically utilized to simulate transonic axial compressor rotors.

We employ certain turbulent quantities within the proposed framework, specifically the turbulence kinetic energy $k$ and the specific dissipation rate $\omega$, for normalization purposes, as delineated in Table \ref{Table: input normalization}. Due to discrepancies between the statistical turbulent quantities computed by DNS and the modeled turbulent quantities derived from RANS, it becomes important to utilize turbulent quantities consistent with the high-fidelity mean flow fields as a normalization factor. We commence by interpolating high-fidelity mean flow data onto the RANS computational grid to acquire compatible turbulent quantities. Subsequently, the interpolated high-fidelity mean flow data are introduced into the $k$ equation and $\omega$ equation. The mean flow is held constant throughout this process, yielding a compatible turbulence flow field, as outlined in the iterative methodology described by Yin et al. \cite{yin_iterative_2022}.

\begin{figure}
\centering \includegraphics[width=135mm]{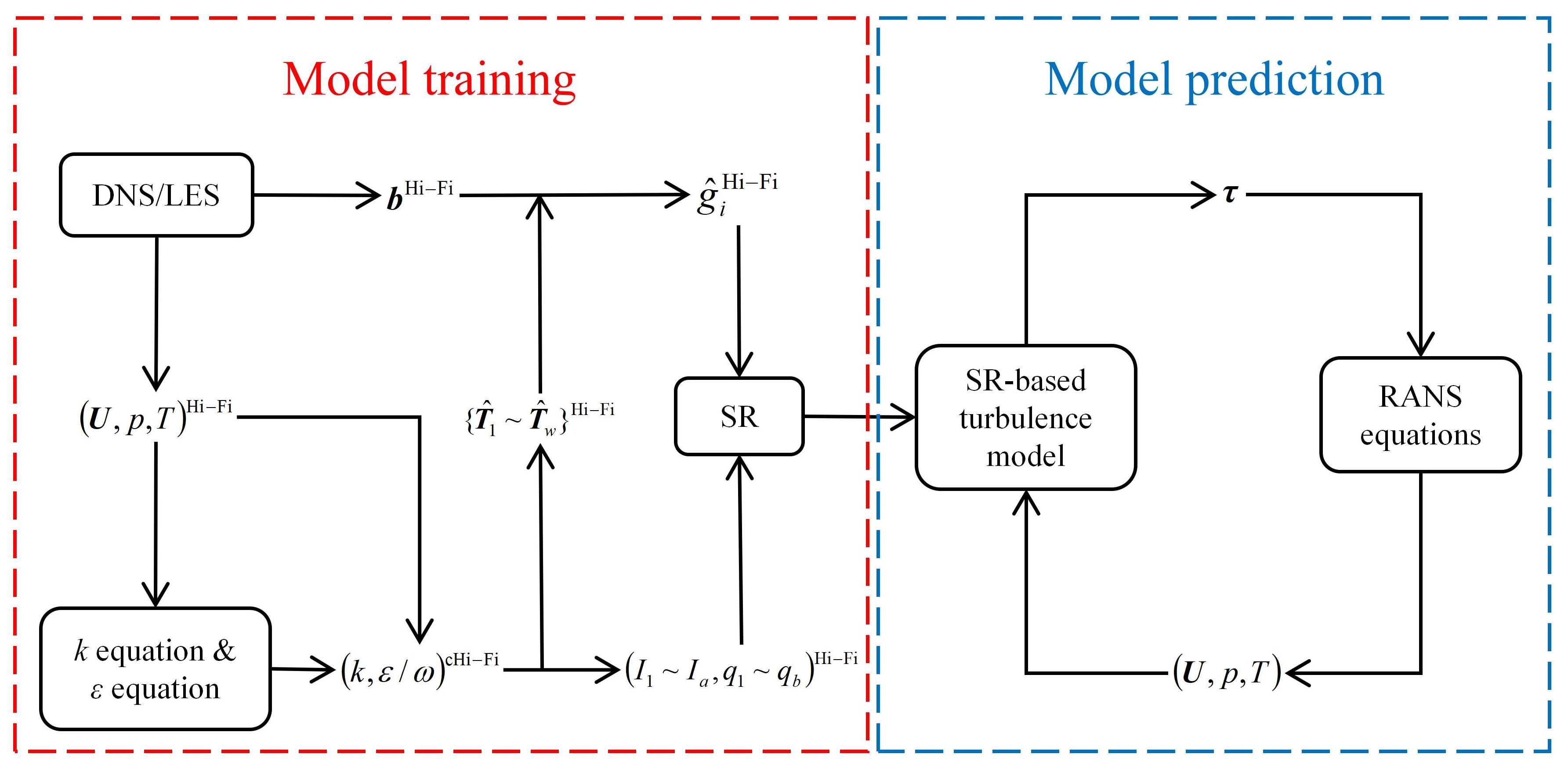} \caption{The framework of our symbolic regression-based turbulence model.}\label{Framework}
\end{figure}

The methodological framework employed in this investigation is elucidated in detail below. Fig. \ref{Framework} provides a comprehensive schematic representation of our proposed approach.

1. Interpolate high-fidelity DNS or LES data onto the RANS mesh to obtain high-fidelity mean flow quantities $(\boldsymbol{U}, p, T)^{\mathrm{Hi-Fi}}$ and Reynolds stress field $\boldsymbol{\tau}^{\mathrm{Hi-Fi}}$.
    
2. Under frozen mean flow conditions, substitute high-fidelity mean flow data $(\boldsymbol{U}, p, T)^{\mathrm{Hi-Fi}}$ into turbulence model scale transport equations ($k$-equation and $\varepsilon$-equation) to generate compatible turbulence fields $(k, \varepsilon / \omega)^{\mathrm{cHi-Fi}}$.
    
3. Construct input features $\left(I_1 \sim I_{\mathrm{a}}, q_1 \sim q_{\mathrm{b}}\right)^{\mathrm{Hi-Fi}}$ and normalized tensor basis $\left\{\hat{\boldsymbol{T}}_1 \sim \hat{\boldsymbol{T}}_w\right\}^{\mathrm{Hi-Fi}}$ based on $(\boldsymbol{U}, p, T)^{\mathrm{Hi-Fi}}$ and $(k, \varepsilon / \omega)^{\mathrm{cHi-Fi}}$. Simultaneously, calculate the normalized tensor basis coefficients $\hat{g}^{\mathrm{Hi-Fi}}$ according to Eq. (\ref{eq: gi_normalized}), using $\boldsymbol{\tau}^{\mathrm{Hi-Fi}}$ and the normalized tensor bases $\left\{\hat{\boldsymbol{T}}_1 \sim \hat{\boldsymbol{T}}_w\right\}^{\mathrm{Hi-Fi}}$.
    
4. Train machine learning algorithms using $\left(I_1 \sim I_{\mathrm{a}}, q_1 \sim q_{\mathrm{b}}\right)^{\mathrm{Hi-Fi}}$ as input and $\hat{g}^{\mathrm{Hi-Fi}}_i$ as output.
    
5. Embed the trained machine learning model into the TCAE code. During the numerical solution iteration process, calculate the Reynolds stress tensor $\boldsymbol{\tau}$ using mean flow variables $(\boldsymbol{U}, p, T)$, and incorporate it into the RANS equation system for a coupled solution until the flow field satisfies convergence criteria.

\subsection{Dataset}
\label{sec:Dataset}

\begin{figure}
\centering \includegraphics[width=130mm]{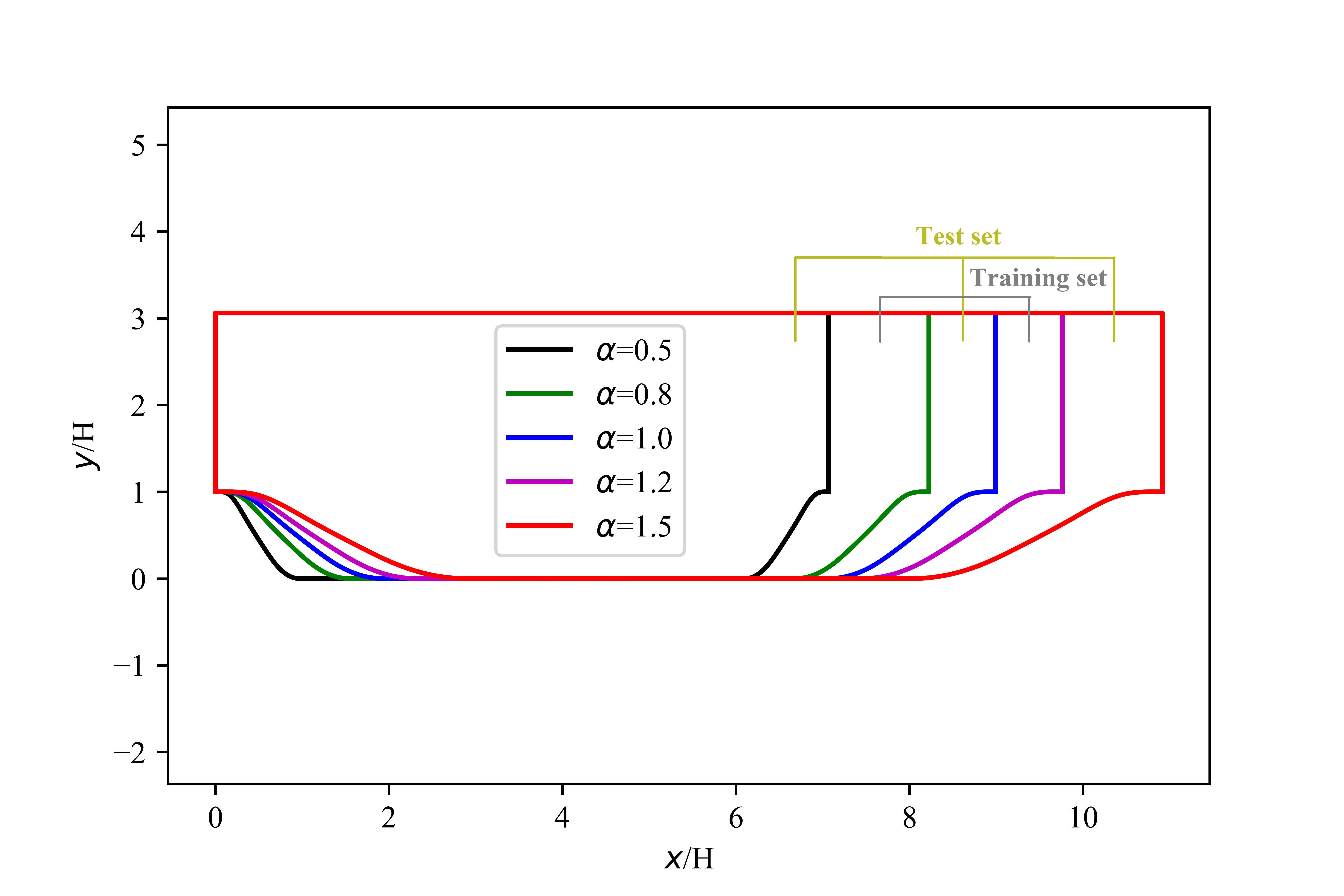} \caption{Parameterized period hill geometries with different $\alpha$.}\label{Parameterized period hill geometries with different}
\end{figure}

In this study, we employ DNS data of parameterized periodic hill flows from Xiao et al. \cite{xiao_flows_2019} as our training dataset. The geometries of these parameterized periodic hills with varying $\alpha$ values are illustrated in Fig. \ref{Parameterized period hill geometries with different}. For our analysis, we utilize data with $\alpha = 0.8$ and $\alpha = 1.2$ as the training set while reserving $\alpha = 0.5$, $1.0$, and $1.5$ for part of the test set. This experimental design allows us to evaluate our model's performance on interpolated ($\alpha = 1.0$) and extrapolated ($\alpha = 0.5$ and $1.5$) flow conditions.

Table \ref{tab:test_cases} presents the flow scenarios for the whole test case. These test cases enable a comprehensive assessment of the generalizability of our symbolic regression-based turbulence model across multiple levels.

\begin{table}
    \centering

    \caption{Test case flow scenarios}
    
    \begin{tabular}{p{2.5cm}p{3.5cm}p{6.5cm}}
        \hline
        Test case flow scenario & Flow Characteristics & Description \\\hline
        
        Periodic hill & 2D, Incompressible, Low Reynolds number, Flow separation induced by bluff body & Separated flow over a bluff body with geometry analogous to the training set. This case evaluates the predictive capability of the symbolic regression model for geometries closely resembling those in the training dataset and can be utilized to assess level 1 generalizability. \\
        
        Zero pressure gradient flat plate & 2D, Incompressible, Low Reynolds number & Evaluate whether the symbolic regression turbulence model adversely affects the baseline model's prediction accuracy for wall-bounded flow quantities. This case can be utilized to assess level 2 generalizability. \\
       
        NACA0012 3D airfoil incompressible flow & 3D, Incompressible, High Reynolds number, Flow separation induced by bluff body & Separated flow over a bluff body featuring a geometry and Reynolds number distinctly differs from those in the training set. This case can be employed to assess level 3 generalizability. \\
        
        NASA Rotor 37 and NASA Rotor 67 transonic axial compressor rotors & 3D, Compressible, Transonic, Separated flow induced by a combination of bluff-body geometry and adverse pressure gradient & Complex engineering flows significantly differ from those in the training set, representing realistic turbomachinery operating conditions. This case assesses the model's level 4 generalizability. \\
        \hline
    \end{tabular}

\label{tab:test_cases}
    
\end{table}
 
\section{Results}\label{sec:Results}

\subsection{Symbolic regression training results}
\label{sec:Symbolic_regression_training_results}

\begin{figure}
\centering \includegraphics[width=130mm]{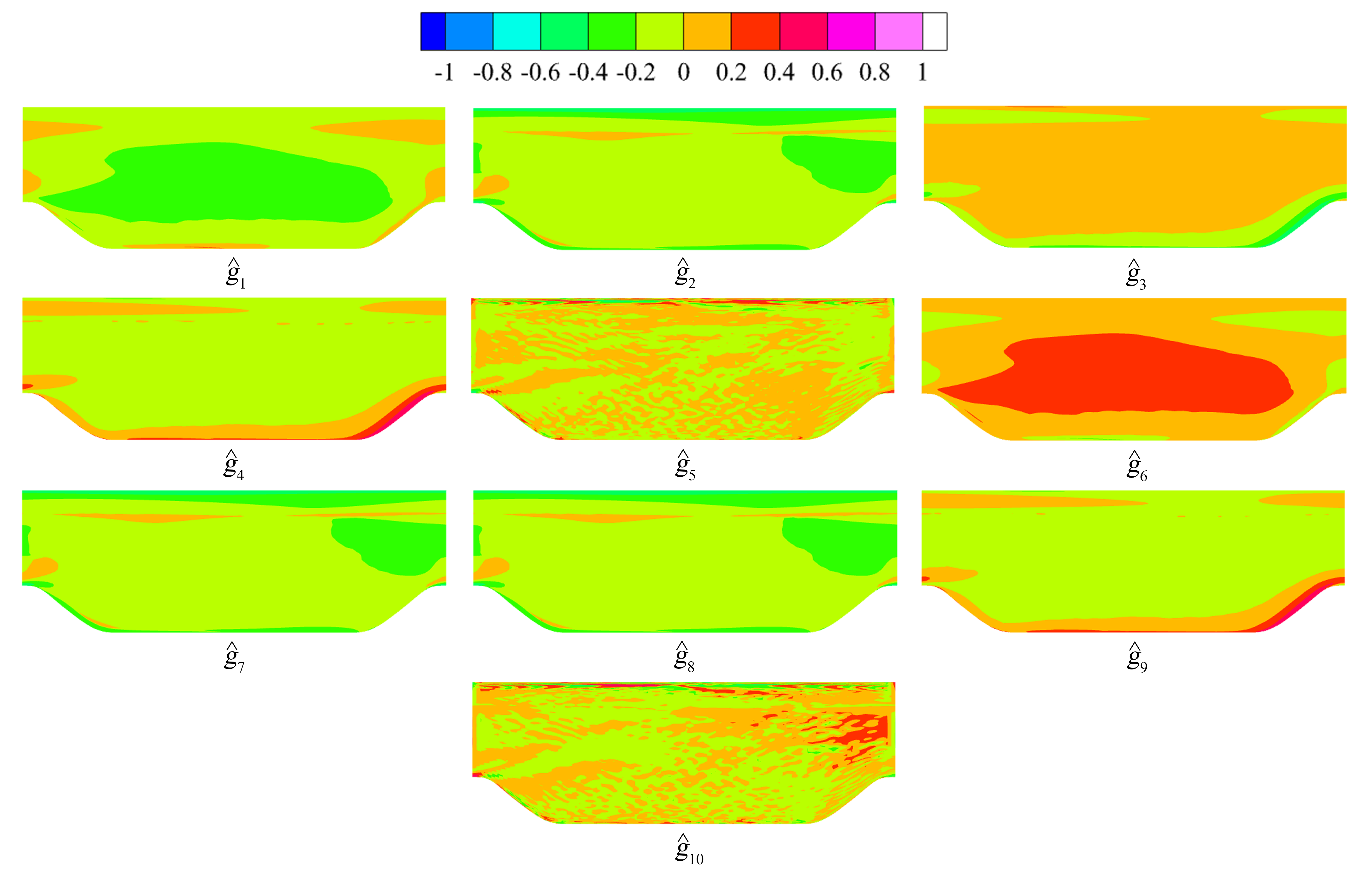} \caption{The contour representation of normalized tensor basis coefficients for the periodic hill configuration with $\alpha = 1.0$.}\label{DNS_ghat}
\end{figure}

Fig. \ref{DNS_ghat} presents a contour representation of the normalized tensor basis coefficients $\hat{g}_i$ for the periodic hill configuration with $\alpha = 1.0$. Notably, all prediction labels are confined within the $[-1, 1]$ range, conforming to the derivation presented in Section \ref{sec:General effective-viscosity hypothesis and its normalized form}. The visualization reveals the sign of each $\hat{g}_i$, indicating whether their corresponding terms contribute positively or negatively to the non-dimensional Reynolds deviatoric tensor $\boldsymbol{b}$. Furthermore, the spatial distributions of $\hat{g}_5$ and $\hat{g}_{10}$ exhibit pronounced discontinuities in this $\alpha = 1.0$ configuration, suggesting that the spatial distributions of $\boldsymbol{\hat{T}_5}$ and $\boldsymbol{\hat{T}_{10}}$ lack smoothness. According to Table \ref{Table: degree}, $\boldsymbol{\hat{T}_5}$ and $\boldsymbol{\hat{T}_{10}}$ possess degrees of 3 and 5, respectively. Based on these observations, we opt to utilize only normalized tensor basis and tensor invariants of degrees 1 and 2, specifically $\{\boldsymbol{\hat{T}_1}, \boldsymbol{\hat{T}_2}, \boldsymbol{\hat{T}_3}, \boldsymbol{\hat{T}_4} \}$ and $\{ I_1, I_3, I_4, I_5, I_{15}, I_{16}, I_{17} \}$.

The symbolic regression results derived through the methodology delineated in Section \ref{sec:Symbolic regression} are presented herein:
\begin{align}
\left\{\begin{array}{llll}
\hat{g}_1 = -0.0833 \left[ q_3 (1 - q_4^2) + q_4^2 q_1 + q_7 (q_4 + 1) \right] \\
\hat{g}_2 = 0.141 q_3 - 0.446 \\
\hat{g}_3 = -0.0432 q_2 \\
\hat{g}_4 = 0.0508 q_2
\end{array}\right.
\label{eq: SR_results}
\end{align}
These four equations' training set loss values are 0.00344, 0.18075, 0.03470, and 0.03676, respectively. Notably, all variables appearing in Eq. (\ref{eq: SR_results}) are derived exclusively from extra features, with none originating from tensor invariants. This empirical finding strongly suggests that extra features possess greater predictive significance than tensor invariants for accurately determining normalized tensor coefficients. The symbolic regression-based turbulence model without normalization results are presented in \ref{sec: Results_of_symbolic_regression-based_turbulence_model_without_normalization}.

\subsection{Periodic hill flows}
\label{sec:Periodic_hill_flows}

The computational domain of periodic hill flows is illustrated in Fig. \ref{Parameterized period hill geometries with different}. The simulations are conducted at Reynolds number $Re = 5600$. The detailed mesh configuration employed in the RANS simulations and DNS results can be found in Xiao et al. \cite{xiao_flows_2019}.

\begin{figure}
\centering \includegraphics[width=135mm]{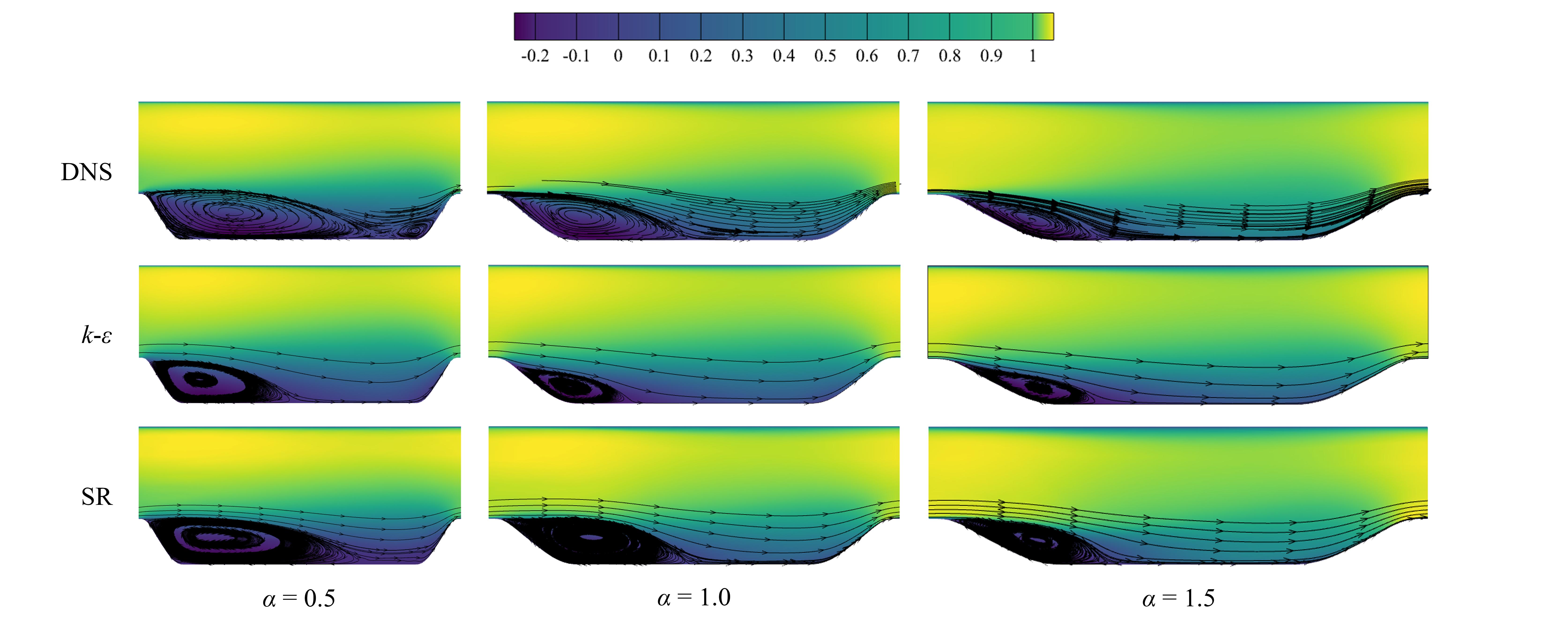} \caption{Streamlines from various simulation results of the test sets. The contours represent $U_x/U_b$.}\label{Streamline}
\end{figure}

The validation dataset for periodic hill flows has been introduced in Section \ref{sec:Dataset}, consisting of cases with $\alpha = 0.5, 1.0, 1.5$ that serve as the validation set for evaluating the performance of our model on periodic hill flow configurations. Fig. \ref{Streamline} illustrates the streamlines derived from various simulation results across the test sets, with contours representing $U_x/U_b$. For the $\alpha = 0.5$ case, the symbolic regression model demonstrates significantly improved prediction of the largest separation vortex in the flow fields compared to the baseline $k$-$\epsilon$ model. However, the symbolic regression and $k$-$\epsilon$ models fail to predict the downstream small vortex accurately. In the $\alpha = 1.0$ case, the symbolic regression model yields superior results relative to the baseline $k$-$\epsilon$ model. Similarly, for the $\alpha = 1.5$ case, the symbolic regression model exhibits enhanced predictive capability compared to the baseline $k$-$\epsilon$ model.

\begin{figure}
\centering 
\begin{minipage}{0.48\textwidth}
  \centering
  \includegraphics[width=\textwidth]{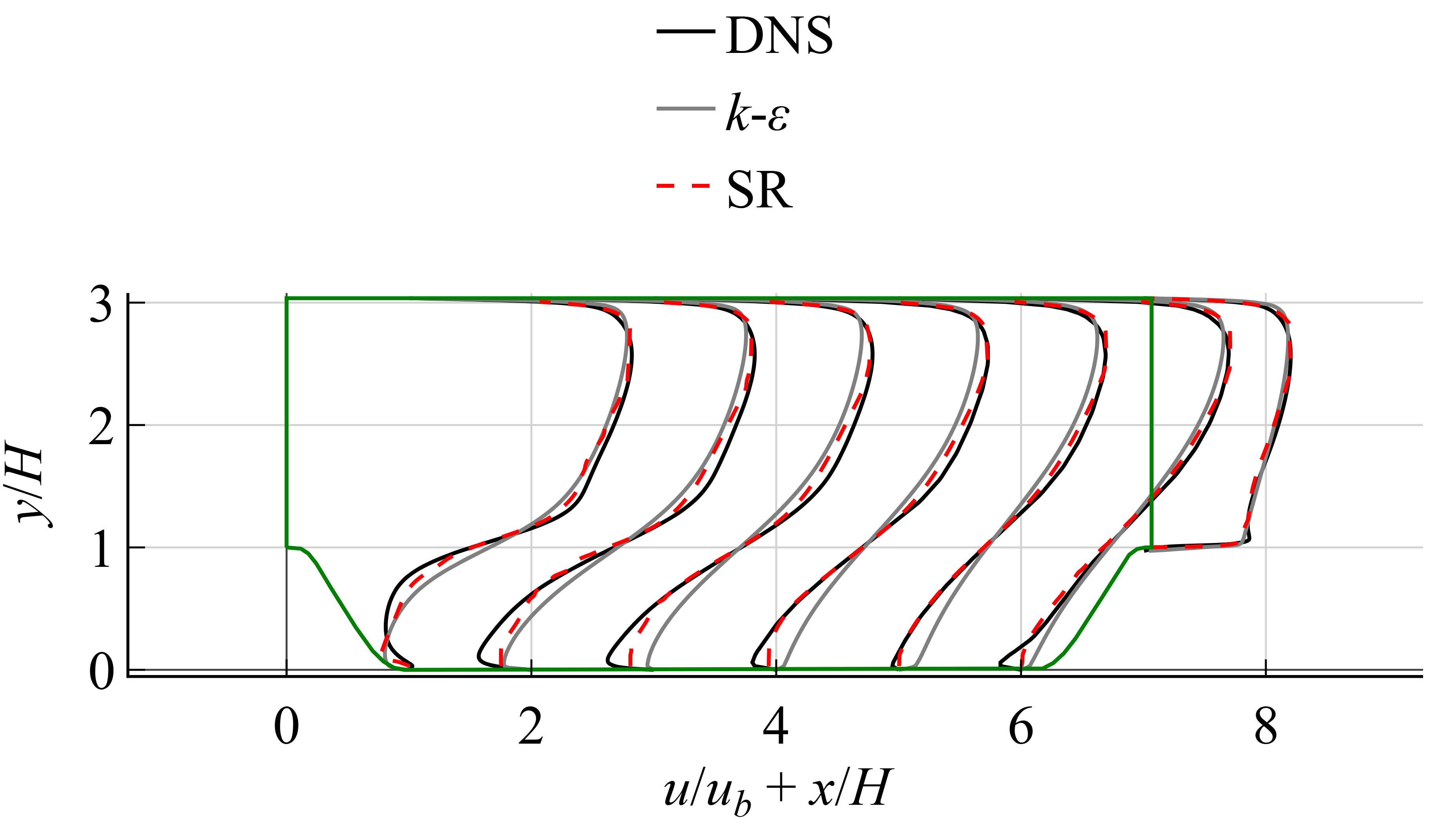}
  \centerline{(a)}
\end{minipage}
\hfill
\begin{minipage}{0.48\textwidth}
  \centering
  \includegraphics[width=\textwidth]{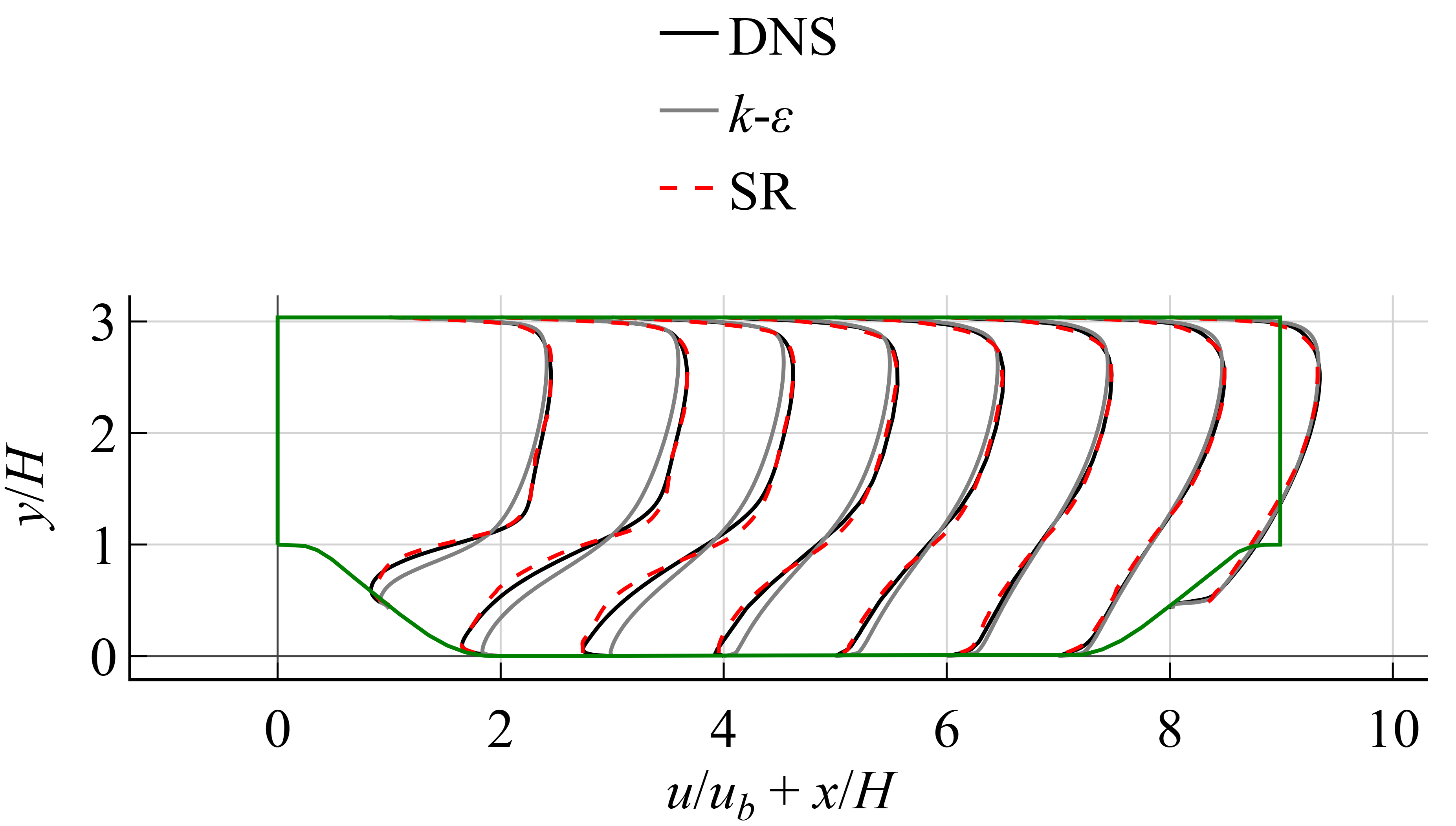}
  \centerline{(b)}
\end{minipage}

\vspace{1em}
\begin{minipage}{0.48\textwidth}
  \centering
  \includegraphics[width=\textwidth]{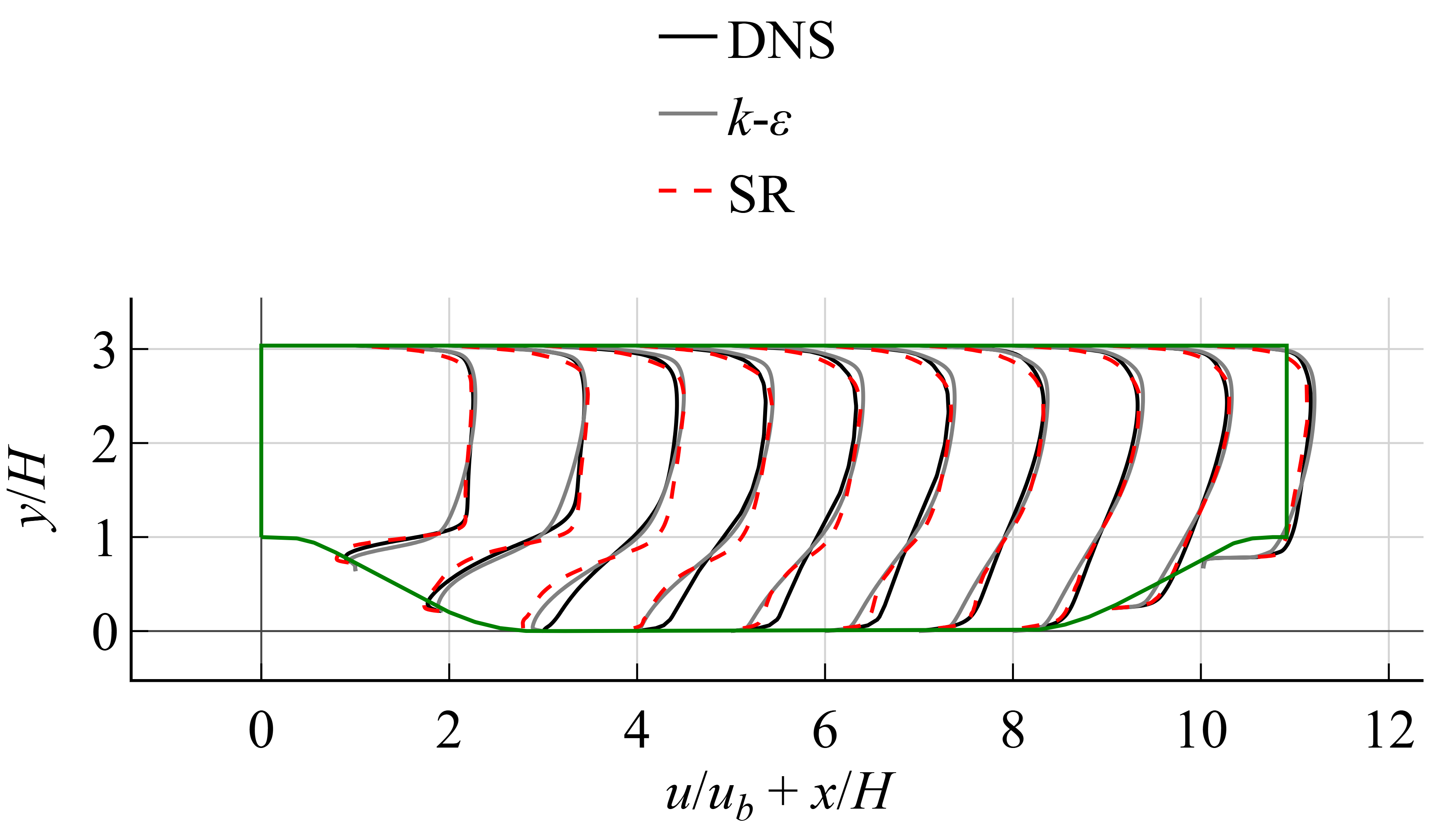}
  \centerline{(c)}
\end{minipage}
\caption{Velocity profiles across the test set from different simulation results. Panels (a), (b), and (c) present the results for the $\alpha = 0.5$, $\alpha = 1.0$, and $\alpha = 1.5$ cases, respectively.}
\label{velocity_profile}
\end{figure}

Fig. \ref{velocity_profile} presents the velocity profiles across the test set obtained from various simulation methodologies. Panels (a), (b), and (c) illustrate the results for the $\alpha = 0.5$, $\alpha = 1.0$, and $\alpha = 1.5$ cases, respectively. The data demonstrate that the symbolic regression model performs better throughout most flow regimes.

\subsection{Zero pressure gradient flat plate flow}
\label{sec:Zero_pressure_gradient_flat_plate_flow}

\begin{figure}
\centering \includegraphics[width=130mm]{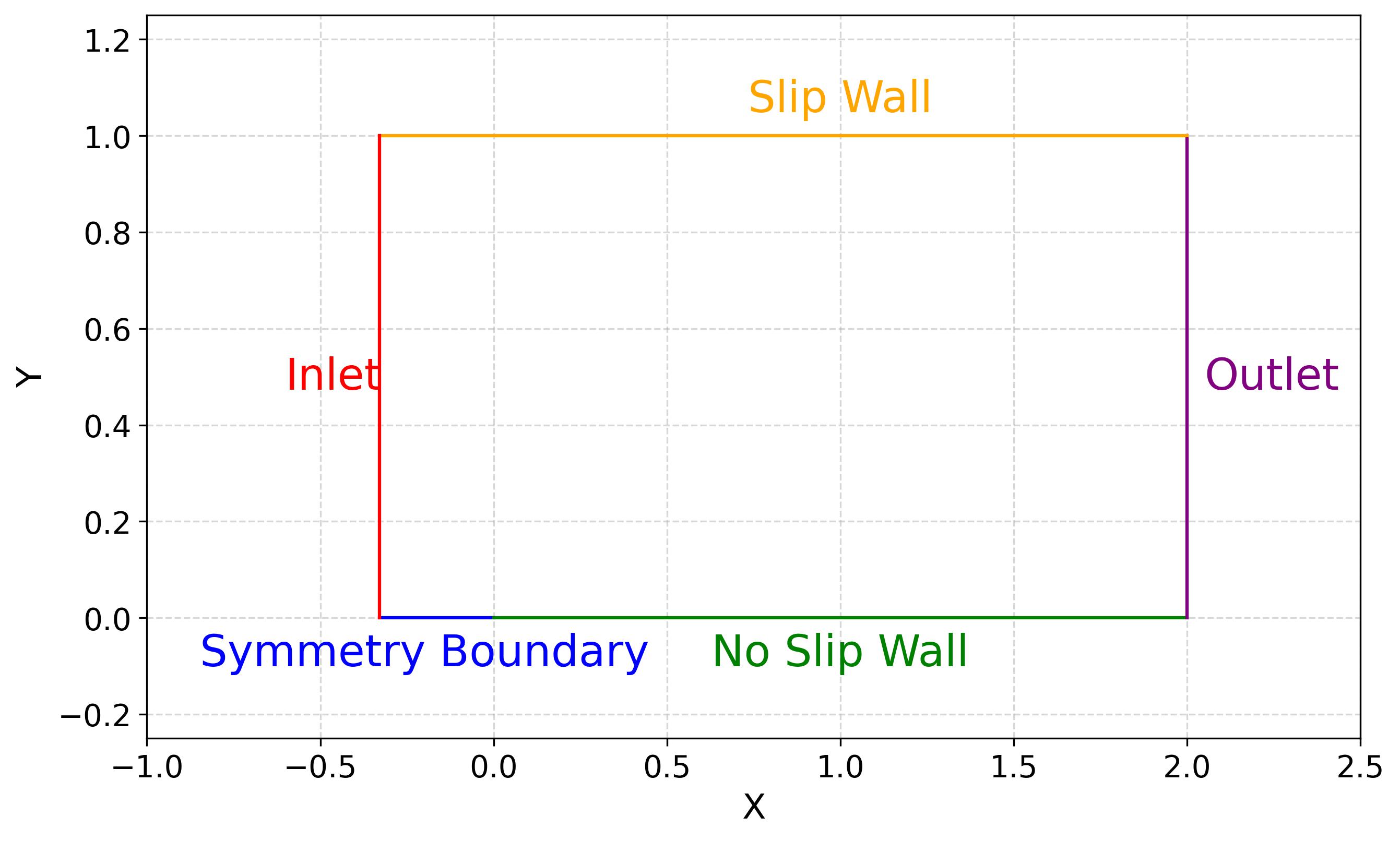} \caption{Computational domain for the zero pressure gradient flat plate flow simulation.}\label{flat_plate}
\end{figure}

The zero pressure gradient flat plate flow is characterized by a Reynolds number, based on the plate length, of $Re = 5 \times 10^6$. The computational domain is illustrated in Fig. \ref{flat_plate}.

\begin{figure}
\centering \includegraphics[width=80mm]{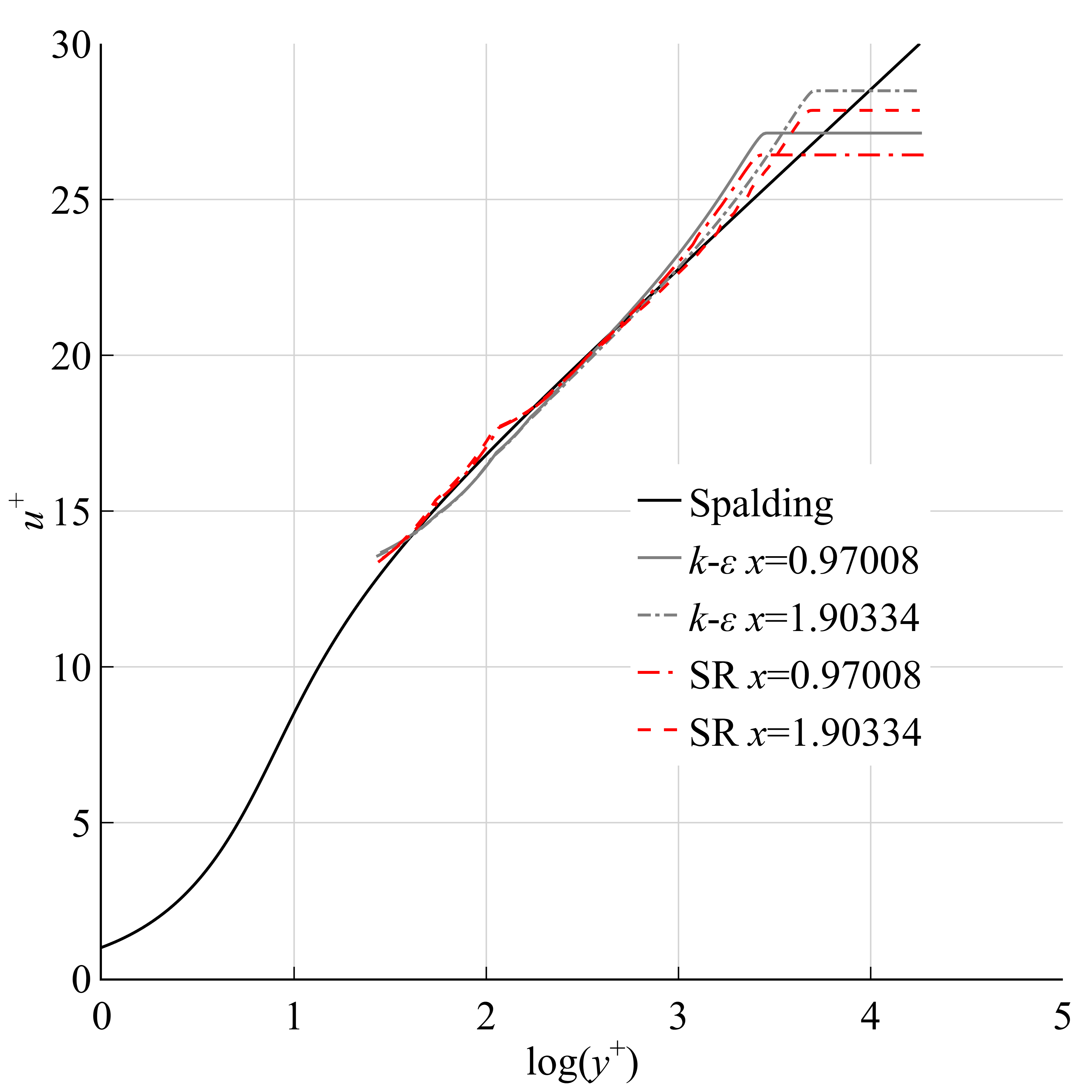} \caption{Comparative analysis of law of the wall results.}\label{law_of_the_wall}
\end{figure}

Fig. \ref{law_of_the_wall} presents a comparative analysis between the computational results obtained from both the baseline $k$-$\epsilon$ model and the symbolic regression model, juxtaposed with the canonical law of the wall formulation proposed by Spalding \cite{spalding_single_1961}. Our model predicts a slight bump near the log($y^+$) = 1 regime, which performs marginally worse than the baseline $k$-$\epsilon$ model. However, our model demonstrates superior performance in other regimes compared to the baseline $k$-$\epsilon$ model. Therefore, we conclude that our model does not significantly compromise the baseline model's prediction accuracy for the law of the wall, and it successfully achieves generalization level 2.

\subsection{Three-dimensional incompressible flow around a NACA0012 airfoil}
\label{sec:Three-dimensional_incompressible_flow_around_a_NACA0012_airfoil}

\begin{figure}
\centering \includegraphics[width=65mm]{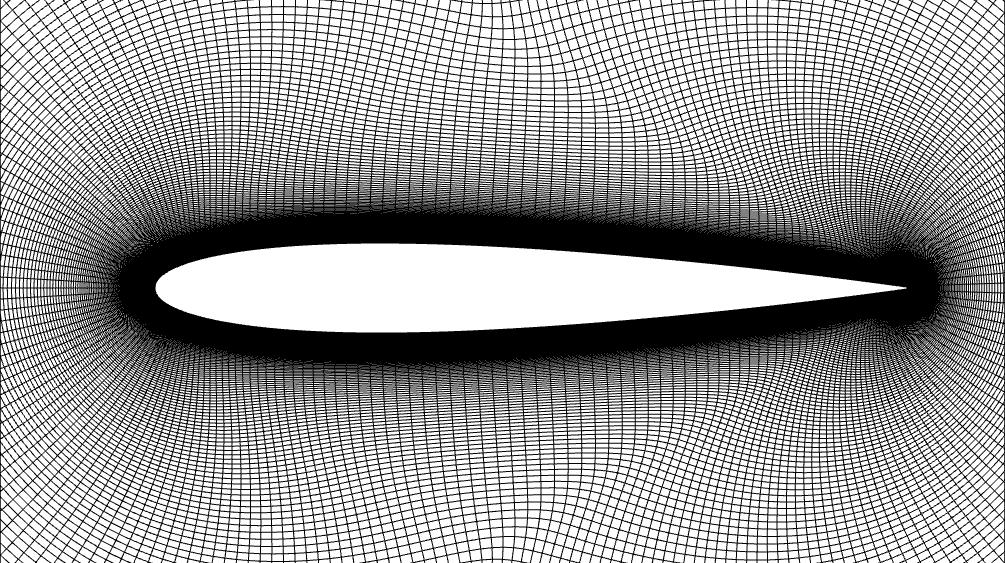} \caption{Computational mesh surrounding the NACA0012 airfoil.}\label{NACA0012_mesh}
\end{figure}

In this section, we evaluate the performance of our symbolic regression-based turbulence models on the three-dimensional incompressible flow around a NACA0012 airfoil. The simulation is conducted at $Re = 3 \times 10^6$. Fig. \ref{NACA0012_mesh} illustrates the computational mesh configuration surrounding the NACA0012 airfoil. The spanwise dimension extends to 0.2 chord lengths and is discretized with 10 grid points. Experimental data for validation are obtained from Ladson et al. \cite{ladson1987pressure}.

\begin{figure}
\centering \includegraphics[width=135mm]{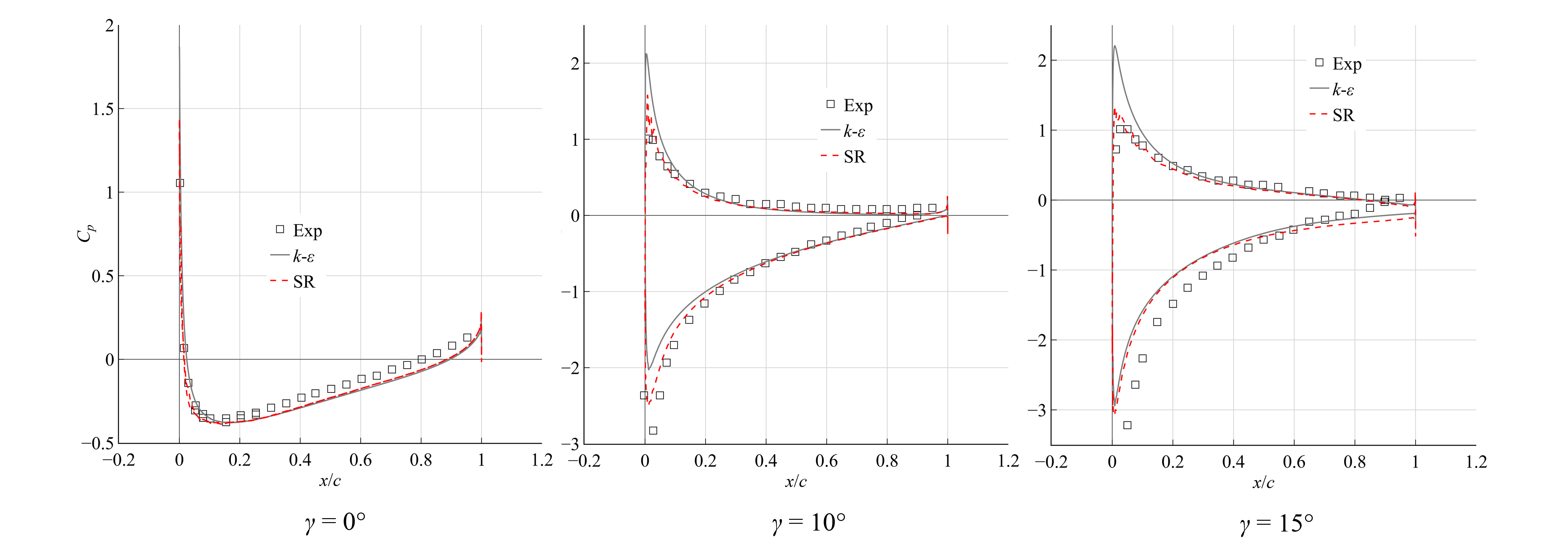} \caption{Comparative analysis of pressure coefficient distributions on NACA0012 airfoil at multiple angles of attack $\gamma$.}\label{NACA0012}
\end{figure}

Fig. \ref{NACA0012} presents a comparative analysis of pressure coefficient distributions along the NACA0012 airfoil at various angles of attack. For the $\gamma = 0^{\circ}$ case, the baseline $k$-$\epsilon$ model and the symbolic regression model yield comparable results. However, for the $\gamma = 10^{\circ}$ and $\gamma = 15^{\circ}$ cases, the symbolic regression model demonstrates enhanced predictive accuracy near the leading edge compared to the baseline $k$-$\epsilon$ model.

\subsection{Transonic axial compressor rotor}
\label{sec:Transonic_axial_compressor_rotor}

\begin{figure}
\centering \includegraphics[width=100mm]{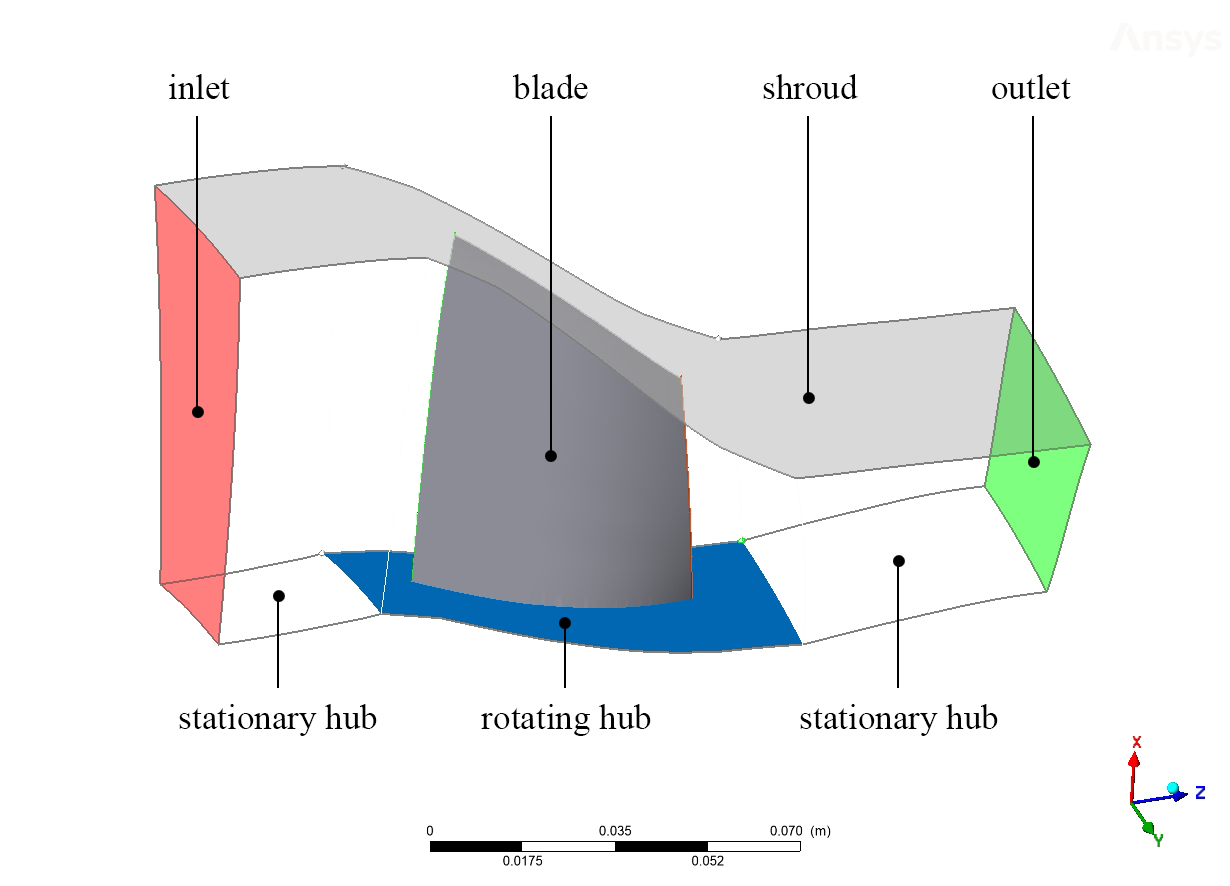} \caption{Three-dimensional computational domain of a single blade of NASA Rotor 37.}\label{Rotor_37}
\end{figure}

In this section, we evaluate the generalizability of our symbolic regression model through analysis of two renowned transonic axial compressor rotors: NASA Rotor 37 and NASA Rotor 67. The design specifications for both rotors are presented in Table \ref{Table: Rotor3767}. These compressor rotors employ shock wave mechanisms to enhance fluid pressure and operate at high rotation speeds (17188.7 rpm for NASA Rotor 37). Both rotors exhibit high total pressure ratios, classifying them as sophisticated engineering-level turbulent flow systems. Consequently, they provide ideal test cases for assessing the level 4 generalizability of our symbolic regression turbulence model. Fig. \ref{Rotor_37} depicts the three-dimensional computational domain for a single blade of NASA Rotor 37. Experimental data for these rotors were obtained from \cite{suder_experimental_1996, strazisar1989laser}. Additional computational methodologies and parameters are documented in our previous publications \citep{ji_tensor_2024, ji_interpreting_2025}.

\begin{table}
    \centering
    
    \caption{Comparison of main design parameters for NASA Rotor 37 and Rotor 67.}
    
    \begin{tabular}{ccc}
    \hline
        Design parameters & NASA Rotor 37 & NASA Rotor 67  \\ \hline
        Rotor total pressure ratio & 2.106 & 1.630  \\ 
        Mass flow rate & 20.188 kg/s & 33.250 kg/s  \\ 
        Rotor wheel speed & 17188.7 rmp & 16043.0 rpm  \\ 
        Number of rotor blades & 36 & 22 \\ \hline
    \end{tabular}
    
    \label{Table: Rotor3767}
\end{table}

\begin{figure}
\centering \mbox{ \subfigure[]{\includegraphics[width=65mm]{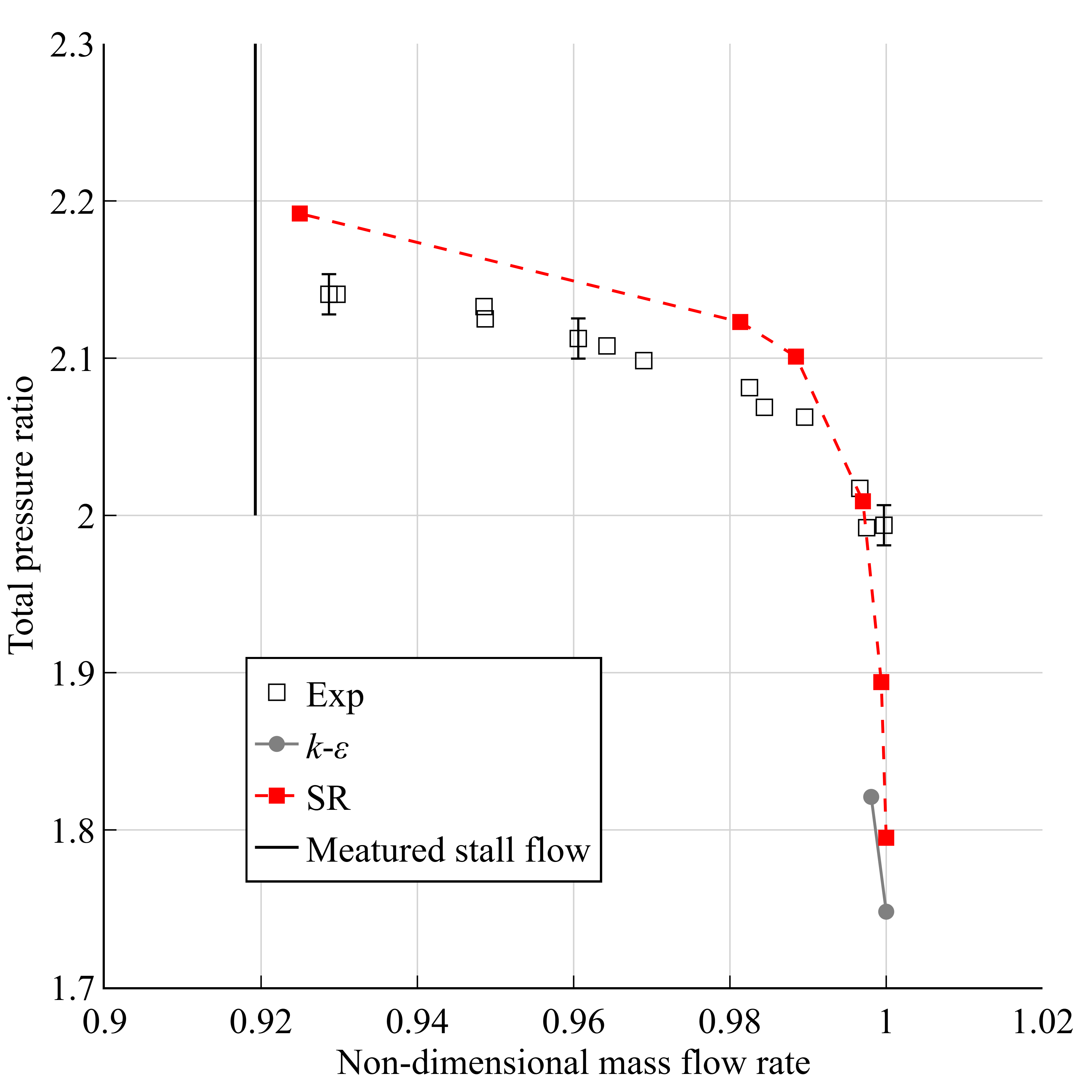}}}\quad \subfigure[]{\includegraphics[width=65mm]{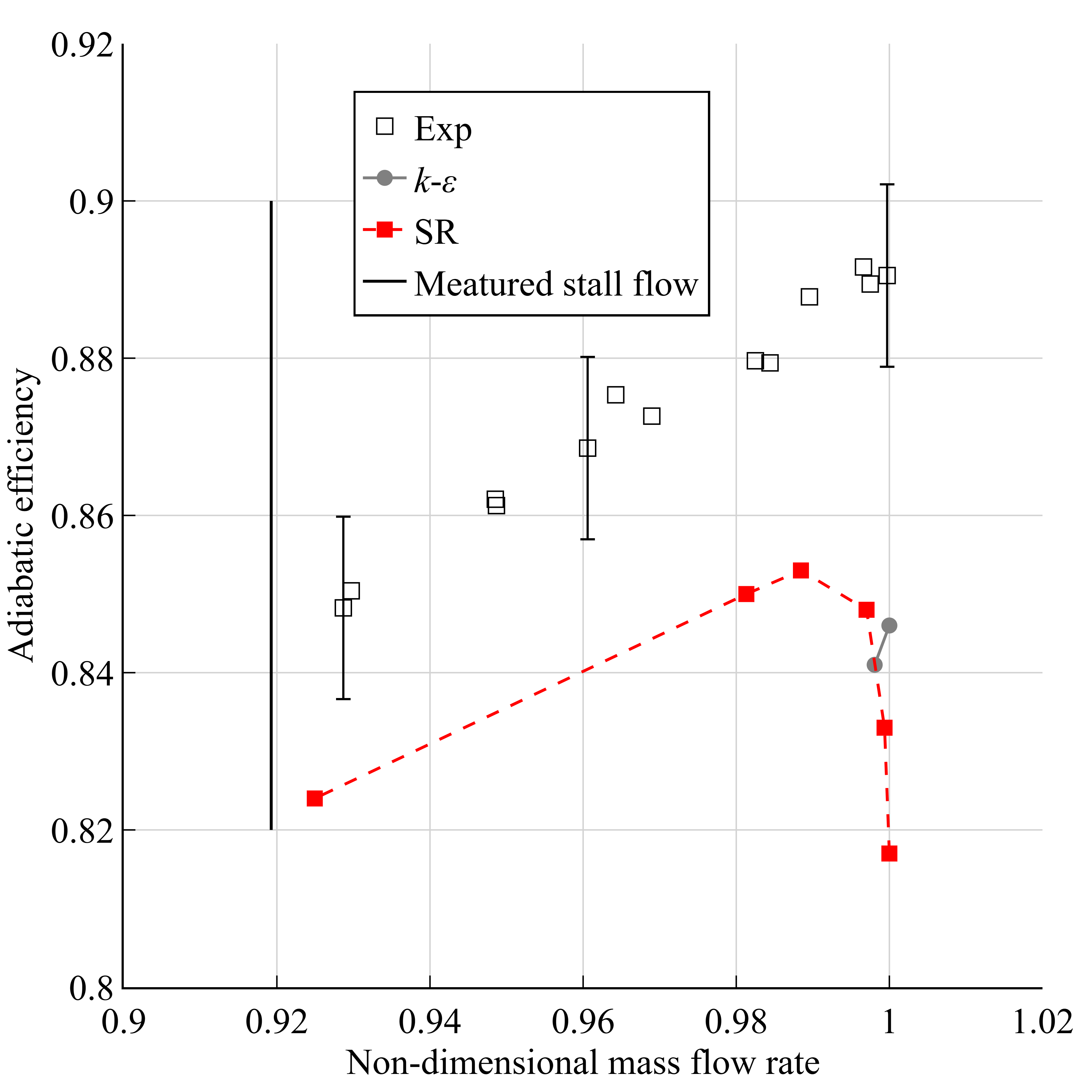}}\quad 
\caption{Overall performance of NASA Rotor 37. Plane (a) represents the total pressure ratio. Plane (b) represents the adiabatic efficiency.}\label{Overall_performance}
\end{figure}

Fig. \ref{Overall_performance} illustrates the overall performance characteristics of the NASA Rotor 37 compressor. Panel (a) depicts the total pressure ratio, while panel (b) presents the adiabatic efficiency. The results demonstrate that the baseline $k$-$\epsilon$ model can only simulate operating conditions near the blockage point. In contrast, the symbolic regression model successfully captures the entire operating range from blockage to stall conditions. This comprehensive coverage demonstrates superior robustness compared to the baseline $k$-$\epsilon$ model. The symbolic regression model provides reasonably accurate predictions for total pressure ratio and adiabatic efficiency across the operational conditions.

\begin{figure}
\centering \mbox{ \subfigure[]{\includegraphics[width=65mm]{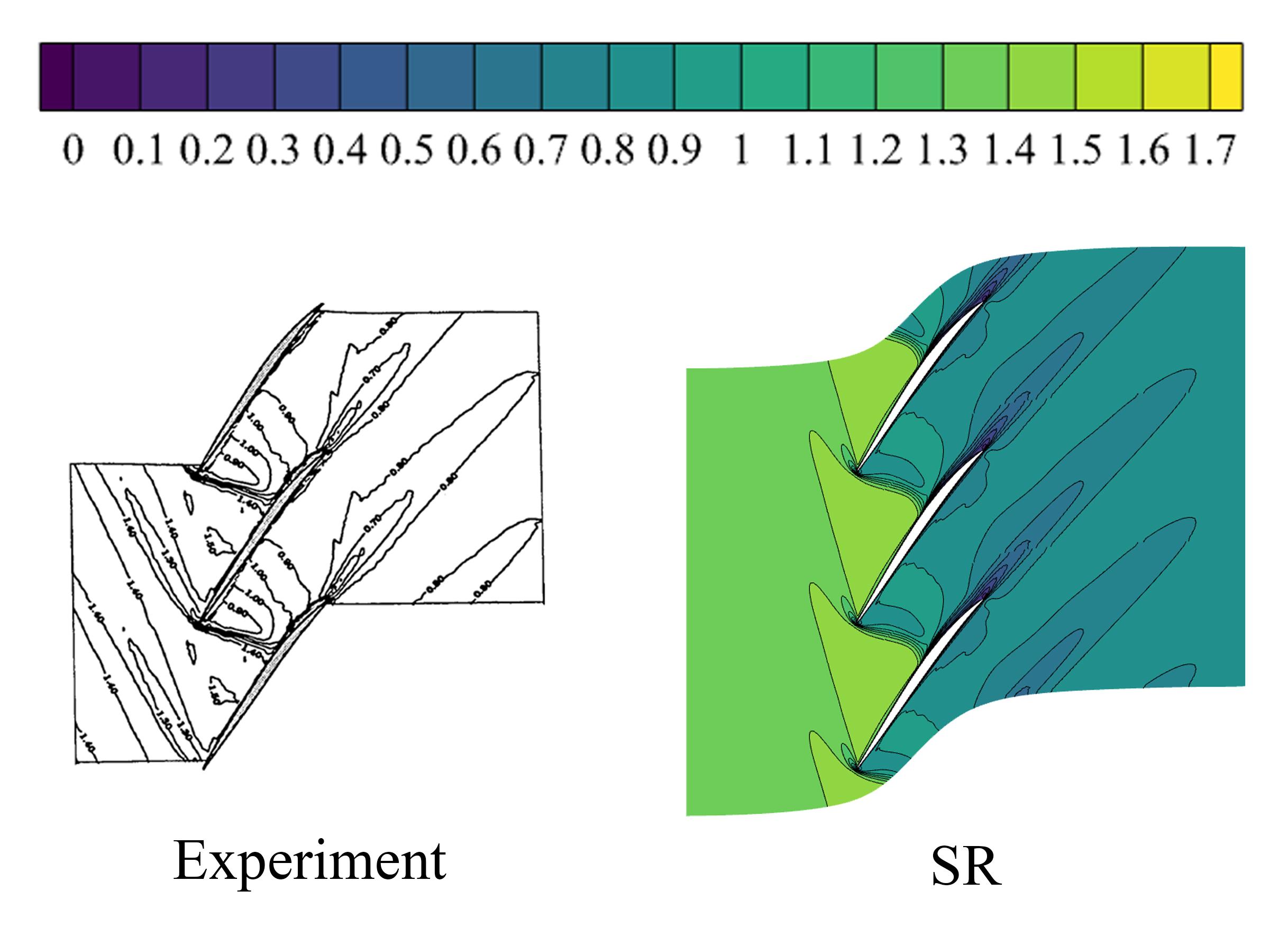}}}\quad \subfigure[]{\includegraphics[width=65mm]{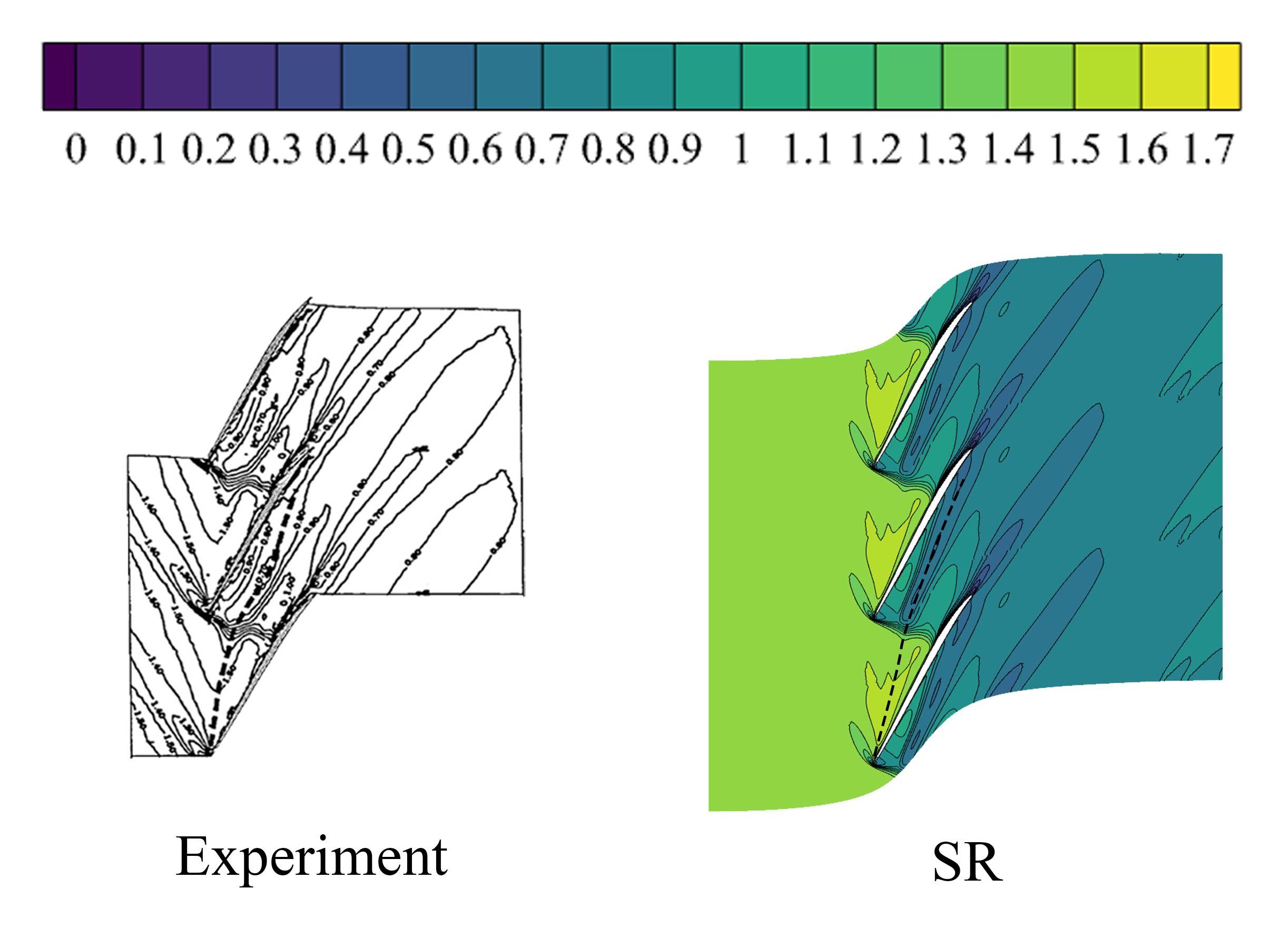}}\quad 
\caption{Relative Mach number contours on 98\% measured chock flow condition of NASA Rotor 37. Plane (a) and Plane (b) illustrate the relative Mach number contours at 70\% span and 95\% span, respectively. The dashed line in Plane (b) delineates the leakage vortex.}\label{Relative_Mach_number}
\end{figure}

Fig. \ref{Relative_Mach_number} illustrates the relative Mach number contours under 98\% measured chock flow conditions. Planes (a) and (b) depict the relative Mach number contours at 70\% and 95\% span, respectively. The dashed line in Plane (b) demarcates the leakage vortex boundary. At 70\% span, our symbolic regression-based turbulence model demonstrates excellent predictive accuracy for shock wave formation and the flow separation induced by shock wave-boundary layer interaction. At 95\% span, the model successfully captures the leakage vortex trajectory and its interaction with the shock wave.

\begin{figure}
\centering \mbox{ \subfigure[]{\includegraphics[width=65mm]{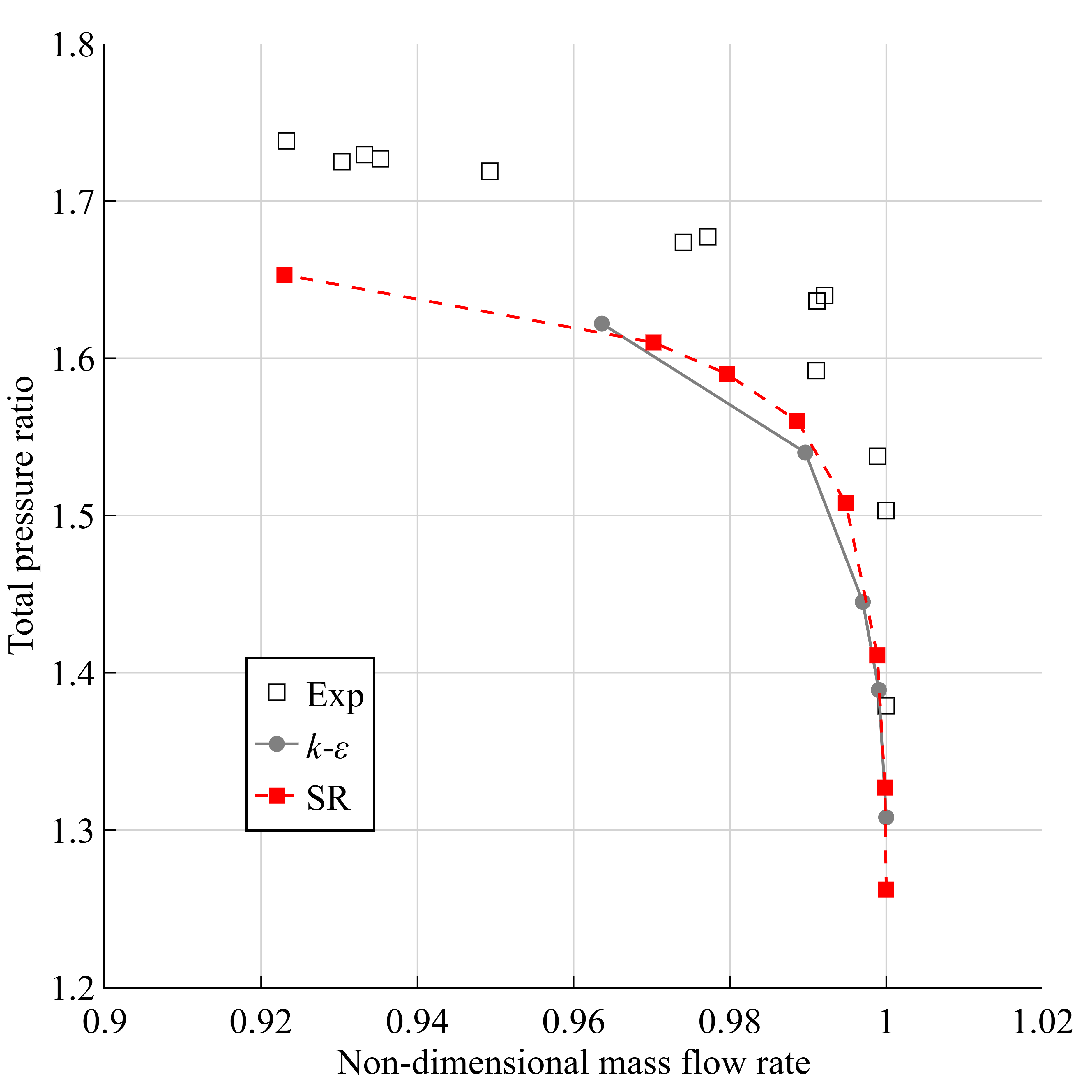}}}\quad \subfigure[]{\includegraphics[width=65mm]{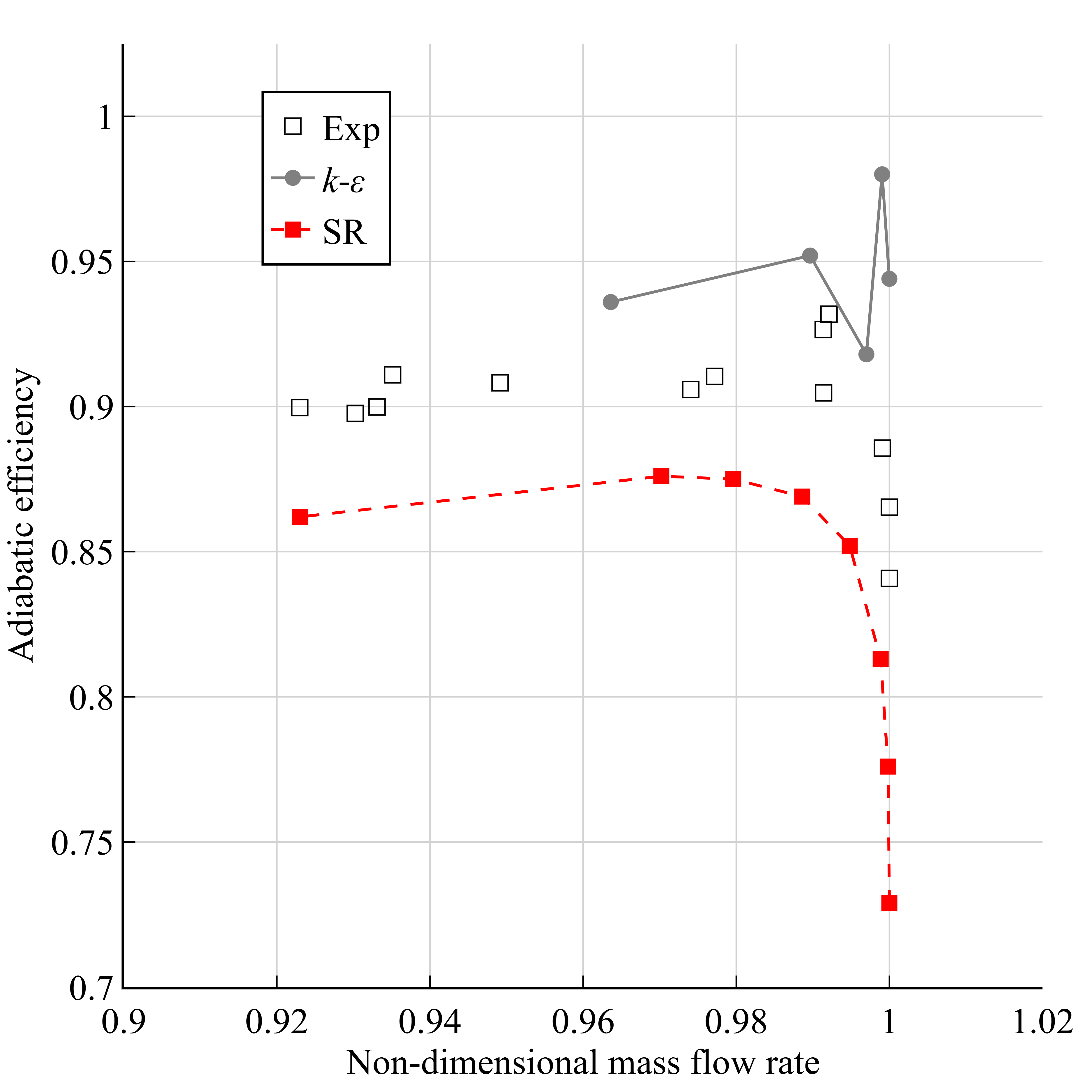}}\quad 
\caption{Overall performance of NASA Rotor 67. Plane (a) represents the total pressure ratio. Plane (b) represents the adiabatic efficiency.}\label{Overall_performance_67}
\end{figure}

Fig. \ref{Overall_performance_67} illustrates the comprehensive performance characteristics of NASA Rotor 67. Panel (a) depicts the total pressure ratio, while panel (b) represents the adiabatic efficiency. The results demonstrate that our symbolic regression model effectively captures the entire operating range from blockage to stall conditions, whereas the baseline $k$-$\varepsilon$ model exhibits significant limitations. Furthermore, our symbolic regression model yields marginally improved simulation accuracy for the total pressure ratio compared to the baseline model. Additionally, the symbolic regression approach generates reasonably accurate predictions of adiabatic efficiency for NASA Rotor 67, further validating the efficacy of the proposed methodology.

It is important to note that although our symbolic regression-based turbulence models demonstrate improvement in simulating transonic axial compressor rotors compared to the baseline $k$-$\varepsilon$ model, certain discrepancies between simulation and experimental results persist. These discrepancies primarily stem from the inherent complexity of engineering flows, where differences between actual experimental conditions and CFD simulation parameters are inevitable. Two main factors contribute to these differences: (a) Inlet Boundary Condition. In this study, we prescribed uniform total pressure and total temperature at the inlet boundary, a simplification that omits detailed profile information. This approximation captures only a portion of the actual experimental conditions documented by Suder \cite{suder_experimental_1996}. For instance, Joo et al. \cite{joo2014large} enhanced boundary condition accuracy by constructing inlet total pressure and total temperature profiles based on experimental velocity data from Suder \cite{suder_experimental_1996}. However, obtaining comprehensive inlet boundary conditions for all operating points remains practically unfeasible. (b) Geometry. The geometry implemented in CFD simulations may not precisely replicate that of the experimental apparatus. The actual geometry is excessively complex to be accurately represented in CFD simulations. In summary, we contend that our symbolic regression-based turbulence models yield reasonably accurate predictions and demonstrate meaningful improvements over the baseline $k$-$\varepsilon$ model, which sufficiently validates the generalizability of our approach.

\section{Discussion}
\label{sec: Discussion}

The features presented in Eq. (\ref{eq: SR_results}) include $q_1$, $q_2$, $q_3$, $q_4$, and $q_7$, which will be discussed in detail in this section.

$q_1$ can be expressed as:
\begin{equation}
\begin{aligned}
q_1=\frac{\frac{1}{2}\left(\|\boldsymbol{R}\|_F^2-\|\boldsymbol{S}\|_F^2\right)}{\left|\frac{1}{2}\left(\|\boldsymbol{R}\|_F^2-\|\boldsymbol{S}\|_F^2\right)\right|+\|\boldsymbol{S}\|_F^2}.
\end{aligned}
\label{eq: q1}
\end{equation}
This parameter represents the ratio of excess mean rotation rate to mean strain rate, commonly called the Q criterion. When $||\boldsymbol{R}||_F^2 \gg ||\boldsymbol{S}||_F^2$, $q_1$ approaches 1, indicating that the mean rotation rate substantially exceeds the mean strain rate. Conversely, when $||\boldsymbol{S}||_F^2 \gg ||\boldsymbol{R}||_F^2$, $q_1$ approaches $-\frac{1}{3}$, signifying that the mean strain rate significantly exceeds the mean rotation rate. Consequently, $q_1 \in [-\frac{1}{3}, 1]$. A positive value of $q_1$ indicates rotation-dominated flow, whereas a negative value of $q_1$ characterizes strain-dominated flow. It is worth noting that when the velocity field is uniform (e.g. when the initial velocity condition is set as a uniform field), $q_1$ may become indeterminate ($\frac{0}{0}$), potentially causing numerical issues in CFD codes. To avoid this issue, we recommend using a non-uniform initial velocity condition, such as the converged results obtained from other turbulence models, as the initial velocity field.

$q_2$ can be expressed as:
\begin{equation}
\begin{aligned}
q_2=\frac{k}{k+\frac{1}{2} U_i U_i}.
\end{aligned}
\label{eq: q2}
\end{equation}
This parameter represents the ratio between turbulent and mean flow kinetic energy. When $k \gg \frac{1}{2} U_i U_i$, $q_2$ approaches 1, indicating that turbulent kinetic energy substantially exceeds mean flow kinetic energy. Conversely, when $\frac{1}{2} U_i U_i \gg k$, $q_2$ approaches 0, signifying that mean flow kinetic energy dominates over turbulent kinetic energy. Consequently, $q_2 \in [0, 1]$. It is important to note that $q_2$ is not invariant concerning coordinate system selection due to the term $\frac{1}{2} U_i U_i$. For instance, when studying turbulent flows on Earth while using the Sun as the coordinate origin, the mean flow velocity must account for the Earth's revolution (orbital motion) and rotation (axial rotation). Therefore, selecting an appropriate coordinate system is essential when implementing the symbolic regression-based turbulence models proposed in this work.

$q_3$ can be expressed as:
\begin{equation}
\begin{aligned}
q_3 = \min \left(\frac{\sqrt{k} d}{50 \nu}, 2\right).
\end{aligned}
\label{eq: q3}
\end{equation}
Kinematic viscosity $\nu$ remains uniform throughout the flow field in most single phase scenarios. In the near-wall region, turbulent kinetic energy $k$ necessarily diminishes to 0 as the distance to the wall approaches zero. Consequently, as the location approaches the wall, $q_3$ tends toward 0. Conversely, as the location moves away from the wall, $q_3$ approaches 2. Therefore, $q_3$ is bounded within the range $[0, 2]$.

$q_4$ can be expressed as:
\begin{equation}
\begin{aligned}
q_4=\frac{U_k \frac{\partial P}{\partial x_k}}{\left|U_k \frac{\partial P}{\partial x_k}\right|+\frac{d^3}{\|\boldsymbol{S}\|_F}}.
\end{aligned}
\label{eq: q4}
\end{equation}
This parameter characterizes the pressure gradient along the streamline and indicates its directionality (positive or negative). It is constrained within the interval $[-1,1]$. As with $q_2$, this parameter exhibits coordinate system dependence, necessitating careful consideration when selecting an appropriate reference frame.

$q_7$ can be expressed as:
\begin{equation}
\begin{aligned}
q_7=\frac{\left|U_i U_j \frac{\partial U_i}{\partial x_j}\right|}{\left|U_i U_j \frac{\partial U_i}{\partial x_j}\right|+\varepsilon}.
\end{aligned}
\label{eq: q7}
\end{equation}
This feature characterizes the ratio of non-orthogonality between velocity and its gradient to the dissipation rate. When $\left|U_i U_j \frac{\partial U_i}{\partial x_j}\right| \gg \varepsilon$, $q_7$ approaches unity, indicating that the non-orthogonality between velocity and its gradient substantially exceeds the dissipation rate. Conversely, when $\varepsilon \gg \left|U_i U_j \frac{\partial U_i}{\partial x_j}\right|$, $q_7$ approaches zero, signifying that the non-orthogonality between velocity and its gradient is negligible compared to the dissipation rate.

Within the theoretical framework of the present study, the Boussinesq hypothesis can be formulated as:
\begin{align}
\left\{\begin{array}{ll}
\boldsymbol{b}_{\text {Bous }}=-\frac{v_t\left(\|\boldsymbol{S}\|_F+\omega\right)}{k}\left\|\boldsymbol{T}_1\right\|_F \hat{\boldsymbol{T}}_1 \\[1em]
\hat{g}_{\text {Bous}, 1}=-\frac{v_t\left(\|\boldsymbol{S}\|_F+\omega\right)}{k}\left\|\boldsymbol{T}_1\right\|_F
\end{array}\right.
\label{eq: Boussinesq_hyphothesis}
\end{align}
Under normal conditions, turbulent kinetic energy $k$, specific dissipation rate $\omega$, and eddy-viscosity $\nu_t$ are strictly positive quantities. Consequently, based on the Boussinesq hypothesis, $\hat{g}_{\text{Bous}, 1}$ invariably assumes negative values. However, the DNS results in Fig. \ref{DNS_ghat} demonstrate that $\hat{g}_1$ can exhibit both positive and negative values. This discrepancy is captured by our symbolic regression model, which predicts that $\hat{g}_1$ can become positive when $q_1$ becomes negative.

The symbolic regression analysis of $\hat{g}_2$, $\hat{g}_3$, and $\hat{g}_4$ reveals that these quantities are expressed as first-degree functions of the supplementary features. Given that $q_3 \in [0,2]$, the symbolic regression model for $\hat{g}_2$ consistently yields negative values, failing to capture the positive values observed in Fig. \ref{DNS_ghat}. Similarly, since $q_2 \in [0,1]$, the symbolic regression representations of $\hat{g}_3$ and $\hat{g}_4$ are constrained to non-positive and non-negative values respectively, which introduces discrepancies when compared to the DNS results illustrated in Fig. \ref{DNS_ghat}.

Despite the impressive performance of our symbolic regression-based turbulence model on the validation set, two aspects of this framework warrant further investigation. First, as demonstrated by the symbolic regression results in Eq. (\ref{eq: SR_results}), extra features exhibit greater significance than tensor invariants. However, additional influential features may have escaped our attention. Therefore, identifying and incorporating other important extra features into this framework remains a critical direction for future work. Second, our implementation of symbolic regression in this study employed exclusively algebraic operators, as listed in Table \ref{Table: Symbolic regression}. More powerful transcendental operators potentially exist that could enhance model performance. It is worth noting that expanding the feature set and operator library would inevitably increase the computational cost of the symbolic regression process.

\section{Conclusion}
\label{sec: Conclusion}

This work proposes a novel tensor basis normalization method to train a symbolic regression-based turbulence model grounded in the general effective-viscosity hypothesis. We use DNS data of periodic hill flows as the training dataset. To evaluate the proposed turbulence model's generalizability, we construct a systematic validation set. The results demonstrate that the symbolic regression-based turbulence model achieves a generalization level 4.

Firstly, we train the symbolic regression-based turbulence model using our novel tensor basis normalization, along with an algorithm we propose to ensure that the symbolic regression results are insensitive to the hyperparameters of the symbolic regression process. Secondly, we construct a systematic validation set, which consists of periodic hill flows with different aspect ratios from the training dataset, zero pressure gradient flat plate flows, three-dimensional incompressible flows over a NACA0012 airfoil, and transonic axial compressor flows from NASA Rotor 37 and Rotor 67. This validation set is designed to evaluate the generalizability of our symbolic regression-based turbulence model. Lastly, we assess the generalizability of the turbulence model using this validation set and find that it achieves a generalization level of 4.

In the future, we plan to explore additional important features and operators to incorporate into our framework for further performance improvement.

\appendix

\section{Results of symbolic regression-based turbulence model without normalization}\label{sec: Results_of_symbolic_regression-based_turbulence_model_without_normalization}

The symbolic regression results, adhering to the rules outlined in Section \ref{sec:Symbolic regression}, are presented below:
\begin{align}
\left\{\begin{array}{llll}
g_1 = (-0.597 - q_4) \\
g_2 = \frac{I_{17}}{I_3} \\
g_3 = \frac{I_5}{I_1} \\
g_4 = \frac{-0.159}{I_3}
\end{array}\right.
\label{eq: SR_results_without_normalization}
\end{align}
These equations have training set loss values of 7.99, 4720, 20800, and $2.58 \times 10^7$, significantly higher than those obtained from the symbolic regression results with normalization. By applying Eq. (\ref{eq: SR_results_without_normalization}) and (\ref{eq: general effective-viscosity hypothesis}), we obtain a symbolic regression-based turbulence model without normalization.

\begin{figure}
\centering \includegraphics[width=135mm]{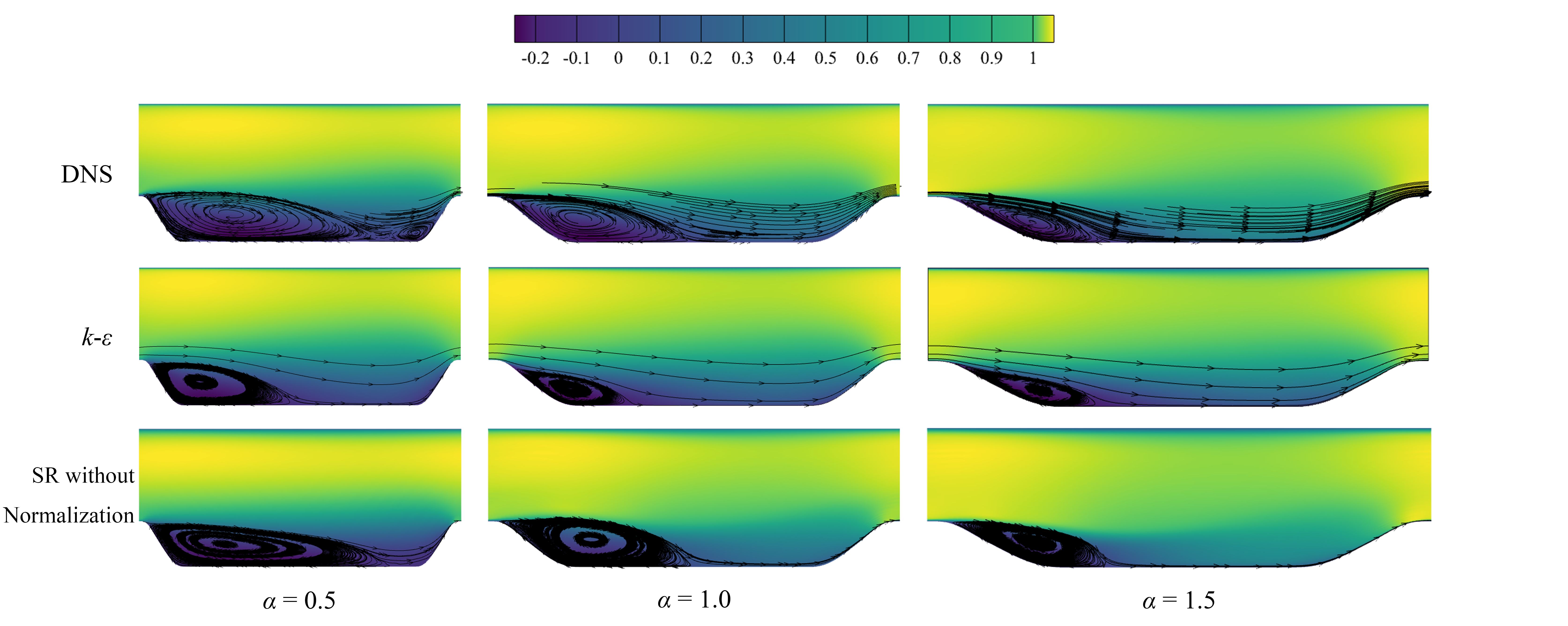} \caption{Streamline results of symbolic regression-based turbulence model without normalization. The contours represent $U_x/U_b$.}\label{Streamline_without}
\end{figure}

Fig. \ref{Streamline_without} illustrates the streamlines predicted by the symbolic regression-based turbulence model without normalization. The contours represent the normalized streamwise velocity ($U_x/U_b$). As demonstrated, this model enhances simulation accuracy for periodic hills across various aspect ratios.

\begin{figure}
\centering \includegraphics[width=80mm]{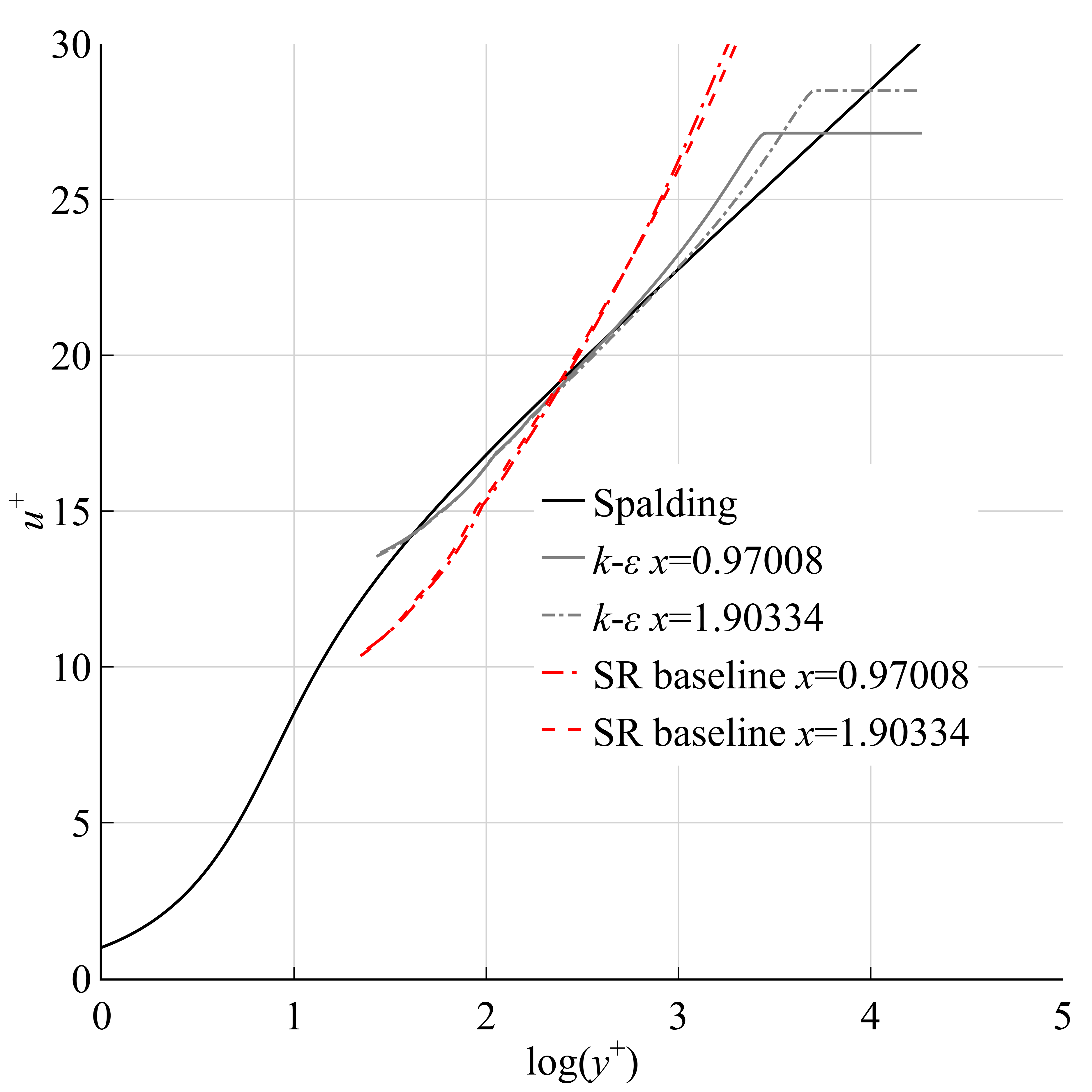} \caption{Law of the wall results predicted by the symbolic regression-based turbulence model without normalization.}\label{law_of_the_wall_without}
\end{figure}

Fig. \ref{law_of_the_wall_without} presents the law of the wall predictions obtained using the symbolic regression-based turbulence model without normalization. The results demonstrate that this model significantly deteriorates the law of the wall simulation performance when compared to the baseline $k$-$\varepsilon$ model. Consequently, the symbolic regression-based turbulence model without normalization fails to attain generalization level 2.

\section*{Acknowledgement}

This research is supported by the National Science and Technology Major Project of China (J2019-II-0005-0025).

\section*{Declaration of Generative AI and AI-assisted technologies in the writing process}

During the preparation of this work the authors used GPT-4 in order to improve readability and language. After using this tool, the authors reviewed and edited the content as needed and take full responsibility for the content of the publication.

\bibliography{cas-refs}
 
\end{document}